\documentclass{ptephy_v1}%%%%%% to generate preprint number with ptep logo

\preprintnumber{XXXX-XXXX} %%% %%% Insert preprint number here

\usepackage{graphicx}
\begin{document}

\title{Thermodynamics of a generalized graphene-motivated $(2+1)$ - dimensional Gross-Neveu model
beyond mean field within the Beth-Uhlenbeck approach}

%%%% To generate auto affiliation numbers please use \author{}\affil{} command
\author{Dietmar Ebert$^1$}

\author{David Blaschke$^{2,3,4}$}
\affil{$^1$Institut f\"ur Physik, Humboldt Universit\"at zu Berlin, Newtonstra{\ss}e 15, 12489 Berlin, Germany \email{debert@physik.hu-berlin.de}}
\affil{$^2$Instytut Fizyki Teoretycznej, Uniwersytet Wroc{\l}awski, pl. M. Borna 9, 50-204 Wroc{\l}aw, Poland
\email{david.blaschke@ift.uni.wroc.pl}}
\affil{$^3$Laboratory of Theoretical Physics, Joint Institute for Nuclear Research, Joliot-Curie str. 6, 141980 Dubna, Russia}
\affil{$^4$National Research Nuclear University (MEPhI), Kashirskoe shosse 31, 115409 Moscow, Russia}
%\affil{$^4$Peoples' Friendship University of Russia (RUDN University), Miklukho-Maklaya str. 6, Moscow, 117198 Moscow, Russia}


\begin{abstract}
We investigate the thermodynamics at finite density of a
generalized $(2 + 1)$-dimensional Gross-Neveu  model of $N$
fermion species with various types of  four-fermion interactions.
The motivation for considering such a generalized schematic model
arises from taking the Fierz-transformation of an effective
Coulomb current-current interaction and  certain symmetry breaking
interaction terms, as considered for graphene-type models in
Ref.~\cite{I.15}. 
We then apply path-integral bosonization
techniques, based on the large $N$ limit, to derive the
thermodynamic potential. 
This includes the leading order mean-field
(saddle point)  contribution as well as the next-order
contribution of  Gaussian fluctuations of  exciton fields. The
main focus of the paper is then the investigation of the
thermodynamic properties of the resulting fermion-exciton
plasma. 
In particular, we derive an extended Beth-Uhlenbeck form
of the thermodynamic potential, discuss the Levinson theorem and
the decomposition of the phase of the exciton correlation into a
resonant and scattering part.
\end{abstract}

\subjectindex{Graphene, 2+1 Gross-Neveu model, Beth-Uhlenbeck equation, exciton correlations}

%\pacs{73.22.Gk, 11.10.Wx, 71.70.Ej}
%\keywords{}

\maketitle

\section{Introduction}

The application of field-theoretic models with local four-fermion
interactions to the investigation of fermion mass generation by
spontaneous breakdown (SB) of  symmetries and related phase
transitions has a long history going back to the famous
BCS-Bogoliubov approach to superconductivity \cite{I.1, I.2}.
Important relativistic generalizations of this approach to strong
interactions considering the dynamical generation of nucleon or
quark masses due to SB of a continuous chiral $\gamma_5$ -
symmetry are the Nambu-Jona-Lasinio (NJL) model \cite{I.3, Eguchi, I.4}
or, in case of lower dimensions $D=(1+1), D=(2+1)$, the
Gross-Neveu (GN$_2$, GN$_3$) model with a discrete or continuous
chiral symmetry \cite{I.5, I.6, I.7, I.8, I.9, I.10}. Generally,
such models are also extremely interesting by the generation of
collective bosonic bound states. A powerful tool for deriving
corresponding effective low-energy theories of collective
bound states (mesons) from relativistic four-fermion models like
NJL quark models are nonperturbative path integral techniques (for
detailed applications and references see, e.g., \cite{I.4}).

It is interesting to note  that in lower-dimensional
condensed-matter systems like polymers \cite{I.11} or graphene
\cite{I.12}, the low-energy fermion excitation spectrum is
effectively described by a (quasi) relativistic  Dirac equation
rather than by a Schr\"odinger equation. Such a similarity with
relativistic QFT was a strong motivation to apply corresponding
path-integral bosonization techniques, developed for relativistic
quark models (\cite{Eguchi}, \cite{I.4}, \cite{I.13}, \cite{I.14}), to a
quasirelativistic planar $(D=2+1)$ system like graphene \cite{I.15}. 
Since the present work  will continue the latter kind
of investigation, let us shortly recall some main ideas of that work. 
There, one starts with an instantaneous four-fermion
interaction obtained by a contact approximation to an
instantaneous Coulomb potential using the "brane-world" or
"reduced QED scenario", proposed in Ref.~\cite{I.16}. 
Next, by performing a Fierz transformation into fermion-antifermion (hole)
$f \bar{f}$- channels yields four different types of four-fermion
interactions. This then leads- by means of the
Hubbard-Stratonovich  trick -to four types of possible exciton
bound state fields $E=\{\sigma_i,\varphi_i\}$, $i=1,2$. As a
result, the four-fermion interactions are now changed into
Yukawa-type bilinear $(E f \bar{f})$-couplings. After this, the
fermion path integral becomes Gaussian and can easily be carried
out. The resulting Fermion determinant, rewritten as  a "Tr ln"
term, then determines the effective low-energy action of
collective exciton fields. Note that for the application of a
suitable $1/N$ expansion technique, it was particular useful to
consider the "multilayer " case of $N=2\widetilde{N}$ degenerate
fermion species (flavors) of real spin $\uparrow$ and
$\downarrow$, living on $\widetilde{N}$ hexagonal monolayers. Thus,
fermions carry here a flavor index $a= (1,...N=2\widetilde{N} )$. It
is particularly important that in case of graphene, there exist
two "valley" d.o.f. of fermions (corresponding to two
"Dirac-points") which combine with two sublattice ("pseudospin")
d.o.f. to form reducible four-spinors of a (quasi) relativistic
Dirac theory. This just allows one to introduce chiral $4\times 4$ Dirac
matrices $\gamma_5$, $\gamma_4$, to consider a new emergent
"pseudospin-valley" chiral $SU(2)_{\rm pv}$ symmetry  and to study both
the dynamical generation of fermion masses by SB of chiral
symmetry and the resulting mass spectrum of composite %bound
excitons.

The considered graphene-motivated GN$_3$ model of Ref.~\cite{I.15}
includes also small symmetry-breaking contributions due to on-site
repulsive lattice interactions (as well as possible
phonon-mediated interactions). Since the corresponding
approximation scheme does not allow a determination of the
effective coupling constants from an underlying microscopic
lattice theory, this generalized GN$_3$ model can evidently be
only considered as a schematic one. Nevertheless, because of its
very general structure, we find it interesting for possible
applications and  specializations.

It is worth mentioning that in distinction to NJL quark models of
strong interactions, the fermions of the considered generalized
graphene-motivated GN$_3$ model are not confined. Thus, contrary
to the forbidden, artifact quark decay of a pion, $\pi\rightarrow
q\bar{q}$, a real decay of a pseudoscalar exciton  $E \rightarrow
f \bar{f}$ is possible.

Finally, let us add some remarks on the problem of
renormalization. As is well-known, the GN$_2$ model is
renormalizable and even asymptotically free \cite{I.5}. On the other
hand, the  GN$_3$ model is perturbatively non-renormalizable, but
becomes renormalizable in the $1/N$ expansion \cite{I.8}. 
Anyway, in the condensed-matter motivated effective GN$_3$ model as
considered in Ref.~\cite{I.15}, there is no real problem with renormalization. 
Indeed, there exists a natural finite momentum
cutoff $\Lambda \sim 1/a$ with $a$ being the lattice spacing.
In graphene $a \approx 2.46 \AA$ and $\Lambda \sim 1/a \approx 5$ keV.

The main goal of the present work is to extend the investigations
of the generalized GN$_3$ model of Ref.~\cite{I.15}, performed at
$T=0$, and to study  its thermodynamics  at finite temperature and
fermion density (chemical potential) beyond the mean-field
approximation. This requires to calculate in the large $N$ limit
the thermodynamic potential of the model including both the
dominating contribution of the fermion loop and the relative order $\mathcal{O}(1/N)$
correction term of exciton fluctuations around mean fields. The
necessary calculations can be performed either by the diagrammatic
summation of Feynman diagrams, as done for $D=(3+1)$-dimensional
NJL models of colored quarks \cite{I.17}, or by powerful path-integral 
methods at finite $T$, $\mu$ \cite{I.18} which might be
further combined with $1/N$ expansion techniques  as in Ref.~\cite{I.10}. 
In the following we prefer to use here an
analogous large-$N$ path-integral approach for calculating the
grand canonical partition function and the related thermodynamic
potential at (relative) order $\mathcal{O}(1/N)$. 
Note that besides explicit symmetry breaking in coupling constants, we allow also for explicit
breaking of chiral symmetry by a bare fermion mass term $m_0$.
This avoids dangerous IR-divergencies \cite{I.10} and, most
importantly, excludes the validity of the Mermin-Wagner-Coleman
(MWC) no-go-theorem \cite{I.19}. 
As is well-known, the latter forbids SB of a continuous symmetry for a $(2+1)$ - dimensional
system at finite temperature.\\

The paper is organized as follows. 
In Sect.~\ref{sec:2} we shortly review the generalized, graphene-motivated $GN_3$ model of
Ref.~\cite{I.15} to be further studied in this paper, including
necessary definitions and motivations. 
Sect.~\ref{sec:3} starts with the path-integral representation of the grand-canonical partition
function and provides via the Hubbard-Stratonovich trick and
after fermion integration the effective bosonized action of
four types of bound-state exciton fields $E=\{\sigma_i,\varphi_i\}$, $i=(1,2)$. 
The path integral over exciton fields becomes
trivial, when evaluated in the large-$N$ (saddle point)
approximation for constant "mean-field" values, and leads to the
mean-field approximation of the thermodynamic potential in subsection \ref{ssec:3a}. 
The respective mean-field values $\bar{\sigma_i},\bar{\varphi_i}$ are
determined from the stationarity (gap) equations as given by the
minima of the thermodynamic potential. In corresponding
subsubsections, we first consider the vacuum gap equations at $T=0$,
$\mu=0$ and then later the gap equation at finite temperature and
chemical potential. 
In Subsect.~\ref{ssec:3b} we calculate the contribution of
exciton fluctuations to the thermodynamic potential at (relative)
order $\mathcal{O}(1/N)$ in Gaussian approximation.
They are given by corresponding determinantal terms of inverse exciton propagators.
The required polarization functions, entering the exciton
propagators, are determined by fermion loop diagrams and
calculated by standard Matsubara summation techniques. 
In Sect.~\ref{sec:4} we present general expressions and equations for the mass spectrum
of exciton bound and resonant states. 
Particular attention is drawn to the possible formation of light (would-be) Goldstone excitons. 
They are generally expected to dominate the thermodynamics of a fermion-exciton plasma at low temperature. 
On the other hand, fermions become light due to restoration of chiral
symmetry at high $T$ and are expected to dominate in that region.

Finally, Sect.~\ref{sec:5} contains the derivation of the celebrated
Beth-Uhlenbeck (BU) formulation for the thermodynamics of a strongly correlated 
fermion system (comp. with \cite{I.17}, \cite{I.20}) which we apply 
for a fermion-exciton plasma considered here.
Note the first application of the BU formula for plasma physics using 
phase shifts for the two-body scattering problem of the Coulomb interaction was given by 
Ebeling \cite{Ebeling}.
A generalization of the BU approach for excitons in dense electron-hole plasmas was 
developed by Zimmermann and Stolz (see \cite{Zimmermann} and references therein)
and taken over to the problem of bound state formation in nuclear matter in Ref.~\cite{Schmidt}.
In the present work we will restrict ourselves to the low-density limit.
Using the pole approximation, we also quote the pressure
of a gas of free excitons and consider a first departure beyond
such an approximation based on a Breit-Wigner type ansatz for the
density of states. Moreover, similarly as in Refs.~\cite{I.17},
\cite{I.21} and \cite{I.22}, we indicate here on the necessity of a
decomposition of the scattering phase shift into a resonant and a
scattering part in order to fulfill the Levinson theorem. 
In Sect.~\ref{sec:6} we summarize our main results and present the conclusions. 
Technical details like the algebra of Dirac matrices
in $D=3$, the explicit solution of the vacuum gap equations and
the calculation of the exciton polarization functions are
relegated to three Appendices.

\section{Generalized GN-model: definitions and motivations}
\label{sec:2}

In the following we shall consider a planar  system of N fermion
species (flavors) described  by a generalized, graphene-motivated
$(2+1)$-dimensional Gross-Neveu (GN)- type of model, analogous to
that considered in Refs.~\cite{I.7,I.15}.

Using euclidean notations, the Lagrangian $\mathcal{L}_e$, 
$\mathcal{L}_e = - \mathcal{L}|_{t\rightarrow-i\tau}$, 
is given  by a sum of two terms
\begin{eqnarray}
\mathcal{L}_e = \mathcal{L}_0 + \mathcal{L}_{\rm int}~,
\label{2.1}
\end{eqnarray}
were $\mathcal{L}_0$ is the free part including a (small) bare fermion mass
$m_0$ and being supplemented by a chemical potential $\mu$,
\begin{eqnarray}
\mathcal{L}_0 =\bar{\psi}\left(\gamma_\mu \partial_\mu - \mu\gamma_3+m_0\right)\psi, 
\label{2.2}
\end{eqnarray}
and $\mathcal{L}_{\rm int}$ describes a generalized local four-fermion
interaction with different coupling structures given by 
\footnote{In this paper we shall use natural units $\hbar = c = k_B = 1$, 
where $k_B$ is the Boltzmann constant. Comparing with the particular case of graphene, the Fermi velocity ${\rm v}_F$
appearing in the effective Lagrangian is then also dimensionless and might be considered here, for notational simplicity, 
as  ${\rm v}_F=1$. The realistic value of ${\rm v}_F$ (${\rm v}_F\approx 1/300$) can be easily reestablished by dimensional 
considerations (see, e.g., Ref.~\cite{I.15}).
Note that for pure graphene $m_0=0$, and fields of fermions $f$ are denoted by $\psi$.}
\begin{eqnarray}
\mathcal{L}_{\rm int} = -\left\{ \frac{G_1}{2N}
\left(\bar{\psi}\psi\right)^2+\frac{G_2}{2N}
\left(\bar{\psi}\gamma_{45}\psi\right)^2 +\frac{H_1}{2N}
\left(\bar{\psi}i\gamma_5\psi\right)^2 +\frac{H_2}{2N}
\left(\bar{\psi}i\gamma_4\psi\right)^2 \right\}~. 
\label{2.3}
\end{eqnarray}

Note that the fermion field $\psi(x)$ in the Lagrangian $\mathcal{L}_e$
transforms as a fundamental multiplet of the flavor group $U(N)$,
i.e. $\psi(x) = \left\{\psi_a(x)\right\}$ with $a = 1,...,N$.
Moreover, each component $\psi_a(x)$  of this multiplet is taken
to be a four-component Dirac spinor that transforms according to a
reducible four-dimensional representation of the rotation group
$SO(3)$.  The reason for this choice is our interest in chiral
symmetry and its possible dynamical breakdown in planar systems
like graphene, which require the existence of a chiral $\gamma_5$
matrix. As is well-known, such a matrix can only be obtained in
$(2 + 1)$ - dimensions by using a reducible $4\times 4$ representation
of Dirac matrices \cite{I.8,I.23}. Note further that $G_i$, $H_i$ are
general coupling constants, and 
$\gamma_\mu$ $\left(\mu =1,...,3\right)$, $\gamma_4$, $\gamma_5$, $\gamma_{45}$ 
are Dirac matrices discussed in detail in Appendix A.

Since the main purpose of this paper is to study the planar
many-body system of fermions and  composite excitons at finite
temperature and fermion density  (finite $\mu$), we find it
convenient to use from the very beginning euclidean notations for
the Lagrangian, coordinates  $x_\mu=(x_1, x_2, x_3 =\tau )=(\vec{x},x_3)$   
$(it =\tau)$ and $\gamma$-matrices. 
Finally, let us remark that spinor and $U(N)$-group indices are not given explicitly and summation
over them is implied everywhere;  moreover, the index at $\mathcal{L}_e$ will now be omitted.

Note that 
%in the chiral limit $m_0 = 0$, 
the kinetic part of the free Lagrangian $\mathcal{L}_0$
%(\ref{2.1})-(\ref{2.3}) 
is invariant under discrete chiral transformations
\begin{eqnarray}
&\gamma_5:&     \psi\rightarrow\gamma_5\psi,~~
\bar{\psi}\rightarrow-\bar{\psi}\gamma_5, \label{2.4}\\
&\gamma_4:&     \psi\rightarrow\gamma_4\psi,~~
\bar{\psi}\rightarrow-\bar{\psi}\gamma_4,,\label{2.5}
\end{eqnarray}
and under the discrete operations of spatial inversion $\mathcal{P}$, time reversal $\mathcal{T}$ and 
charge conjugation $\mathcal{C}$.
Recall that in the reducible (four-dimensional) representation of the Clifford algebra, the definition of the discrete transformations $\mathcal{P, T, C}$ is not unique.
Bearing in mind the condensed-matter roots of the graphene-motivated GN$_3$ model considered above, the discrete symmetry operations will suitably be taken here in such a way that they are consistent with the underlying tight-binding model on the honeycomb lattice (see Ref.~\cite{I.29}). 

By slightly modifying the corresponding definitions of Ref.~\cite{I.29} when using the imaginary time formalism of equilibrium thermodynamics, euclidean notations for Dirac matrices and Grassmann fields for fermions as required in path integrals (for explanations and details, see App. \ref{app:symmetries}), one has

\begin{eqnarray}
\mathcal{P}&:& ~~\psi(\vec{x}, x_3) \rightarrow \gamma_3 \psi(-\vec{x},x_3),~
\bar{\psi}(\vec{x}, x_3) \rightarrow \psi^\dagger(-\vec{x},x_3), \label{2.6a}
\\
\mathcal{T}&:& ~~\psi(\vec{x}, x_3) \rightarrow (i\sigma_2)i\gamma_1\gamma_5 \psi(\vec{x},x_3),~
\bar{\psi}(\vec{x}, x_3) \rightarrow -\bar{\psi}(\vec{x},x_3)i\gamma_1\gamma_5(i\sigma_2), \label{2.6b}
\\
\mathcal{C}&:& ~~\psi(\vec{x}, x_3) \rightarrow i\gamma_1 \bar{\psi}^t(\vec{x},x_3),~
\bar{\psi}(\vec{x}, x_3) \rightarrow (-i\gamma_1\psi(\vec{x},x_3))^t. \label{2.6c}
\end{eqnarray}

Here the spin matrix $\sigma_2$ in $\mathcal{T}$ acts on the indices of physical spin included in the combined 
flavor index $a$ suppressed here, and the superscript $t$ denotes transposition.
(Note that in euclidean space $\mathcal{T}$ does not change the sign of euclidean time $x_3$.)

The respective transformation properties of various fermion bilinears under discrete operations  $\mathcal{P, T, C}$ 
of spatial inversion, time reversal, charge conjugation and under chiral transformations with $\gamma_5$, $\gamma_4$
are shown in table~\ref{tab:1}.
It then follows that $\mathcal{L}_0$ is invariant under $\mathcal{P, T}$ and $\mathcal{C}$ (with the restriction $\mu \to -\mu$ for
$\mathcal{C}$). 
Moreover, since $\mathcal{L}_{\rm int}$ depends quadratically on fermion bilinears, it follows from table~\ref{tab:1} that also the total Lagrangian $\mathcal{L}$ is invariant.

\begin{table}
    \begin{tabular}{|c||cccc||c|}
        \hline
        $\bar{\psi} \Gamma_A \psi$ &
        $\bar{\psi} \psi$ &
        $ \bar{\psi} \gamma_{45} \psi$ &
        $\bar{\psi} i \gamma_{5} \psi $ &
        $\bar{\psi} i \gamma_{4} \psi$ &
        $\bar{\psi} \gamma_{\mu} \psi$
        \\
        \hline\hline
        $\mathcal{P}$ &1&1&-1&-1 & $\bar{\psi} \widetilde{\gamma_{\mu}} \psi$\\
        $\mathcal{T}$ &1&-1&1&1 & $\bar{\psi} \gamma_{\mu} \psi$\\
        $\mathcal{C}$ &1&1&-1&1 & $-\bar{\psi} \gamma_{\mu} \psi$\\
        $\gamma_{5}$ &-1&1&-1&1 & $\bar{\psi} \gamma_{\mu} \psi$\\
        $\gamma_{4}$ &-1&1&1&-1 & $\bar{\psi} \gamma_{\mu} \psi$\\
        \hline
    \end{tabular}
\caption{Transformation properties of various fermion bilinears $\bar{\psi} \Gamma_{A} \psi$, 
$\Gamma_A=(\Gamma_i, \widetilde{\Gamma_i})$ (compare Eq.~(\ref{3.5}) below), 
and the current $\bar{\psi} \gamma_{\mu} \psi$
under the discrete transformations described in the text.
In case of spatial reflections the transformed bilinears depend on reversed coordinates, and we have
$\widetilde{\gamma_\mu}=(-\gamma_1,-\gamma_2,\gamma_3)$.
Note that in euclidean space, spatial components of the current do not change under $\mathcal{T}$.
\label{tab:1}}
\end{table}

Let us next consider also continuous chiral transformations $\gamma_5$ and $\gamma_4$ 
\begin{eqnarray}
&U_{\gamma_5}(1):&     \psi\rightarrow
\exp(i\alpha\gamma_5)\psi,~~
\bar{\psi}\rightarrow \bar{\psi}\exp(i\alpha\gamma_5), \label{2.7}\\
&U_{\gamma_4}(1):&     \psi\rightarrow
\exp(i\tilde{\alpha}\gamma_4)\psi,~~ \bar{\psi}\rightarrow
\bar{\psi}\exp(i\tilde{\alpha}\gamma_4).\label{2.8}
\end{eqnarray}

Taking certain symmetry restrictions like $G_1=H_1$ ($G_2$, $H_2$ arbitrary) or 
$G_1=H_2$ ($G_2$, $H_1$ arbitrary), the Lagrangian $\mathcal{L}$ of Eq.~(\ref{2.1})
is even invariant under continuous chiral  $\gamma_5$ and $\gamma_4$ transformations. 

It is worth remarking that in the case of a continuous
chiral transformation ${\rm U}_{\gamma_5}(1)$ 
%(see Eq.~(\ref{2.7}); $G_1 = G_2 =H_1 = H_2$),
fermion bilinears transform as follows, 
\begin{eqnarray}
 \label{5.12}
% \psi &\rightarrow& \exp(i\alpha \gamma_5)\psi~,\nonumber\\
%\bar{\psi} &\rightarrow& \bar{\psi} \exp(i\alpha \gamma_5)~,\nonumber\\
(\bar{\psi}\psi) &\rightarrow& (\bar{\psi}\psi) \cos(2\alpha) + (\bar{\psi}i \gamma_5 \psi) \sin(2\alpha)~,\nonumber\\
(\bar{\psi}i \gamma_5\psi) &\rightarrow& -(\bar{\psi}\psi) \sin(2\alpha) + (\bar{\psi}i \gamma_5 \psi) \cos(2\alpha)~,\nonumber\\
(\bar{\psi}i \gamma_4\psi) &\rightarrow& (\bar{\psi}i \gamma_4\psi)~,\nonumber\\ 
(\bar{\psi} \gamma_{45} \psi) &\rightarrow& (\bar{\psi} \gamma_{45} \psi)~.
\end{eqnarray}
Analogous relations hold for continuous chiral $\gamma_4$ transformations.\\

Let us remark  that particular cases of the interaction  part
(\ref{2.3}), like e.g. a discrete GN-model $(G_2 = H_{1,2} = 0)$
(see, e.g., Refs.~\cite{I.5}, \cite{I.6}, \cite{I.10}, \cite{I.23}) or
a chiral Nambu-Jona-Lasinio type of model $(G_1=H_1, ~G_2 = H_2 =0)$ 
(see, e.g., Refs.~\cite{I.8, I.10} and references therein) have already been
studied earlier in the literature. Our motivation for studying the
generalized GN-type of model (\ref{2.3}) comes from the
investigation of effective Coulomb interactions in hexagonal,
graphene-like lattice sheets \cite{I.15},
\begin{eqnarray}
\Delta \mathcal{L}^{\rm Coul}_{\rm int} =
\frac{G^C}{2}\left(\bar{\psi}\gamma_3\psi\right)^2, \label{2.9}
\end{eqnarray}
possibly supplemented by a small symmetry-breaking on-side
repulsive interaction term  and analogous terms from
phonon-mediated interactions, which in summary are given by 
\begin{eqnarray}
\Delta \mathcal{L}^{\rm sb}_{\rm int} =
-\frac{\widetilde{G}}{2}\left(\bar{\psi}\psi\right)^2,~\tilde{G}<0~. 
\label{2.10}
\end{eqnarray}
It is worth remarking that in the chiral limit $(m_0 = 0)$ the
free Lagrangian (\ref{2.2}) and the Coulomb-interaction part
(\ref{2.9}) of graphene are invariant under the  emergent
"pseudospin-valley" symmetry group $SU(2)_{\rm pv}$ with generators 
$t_1=\frac{1}{2}\gamma_4$, 
$t_2 =\frac{1}{2}\gamma_5$, 
$t_3 =\frac{1}{2}\gamma_{45}$ 
(see Eqs.~(\ref{A5}), (\ref{A6}) in Appendix A) .
Note that we have introduced the flavor group $U(N)$ for later use of a $1/N$ expansion.
In the corresponding "multilayer" case of graphene (see Ref.~\cite{I.15}), one has
$N=2 \widetilde{N}$ degenerate fermion species (flavors) of real spin $\uparrow$ and $\downarrow$
living on $\widetilde{N}$ hexagonal monolayers.
In addition to the emergent $U(2)_{\rm pv}=U(1)_{\rm pv}\otimes SU(2)_{\rm pv}$ invariance
and the flavor $U(N)$ invariance, then also arises invariance under the larger group $U(2N)$
\footnote{Note that the symmetry-breaking term Eq.~(\ref{2.10}) breaks the larger group $U(2N)$, 
$U(2N) \to U(N)_{t_0}\otimes U(N)_{t_3}$. Here the Lie algebra of the groups $U(N)_{t_0,t_3}$
is given by the direct product of respective generators $t_0=\frac{1}{2}I$, $t_3$ of $U(2)_{\rm pv}$
with the generators of $U(N)$\cite{I.15}.}. 

Next, by performing a Fierz transformation into bound-state
interaction channels for composite exciton fields 
$\Phi_A \sim\bar{\psi}\Gamma_A \psi$, 
$\Gamma_A=\left\{I, \gamma_{45}; i\gamma_5, i\gamma_4\right\}$, 
and discarding, for simplicity, vector/axial-vector type interactions, we obtain from
Eqs.~(\ref{2.9}),(\ref{2.10}) the expression 
$\mathcal{L}_{\rm int} = \Delta \mathcal{L}^{\rm Coul}_{\rm int}+\Delta \mathcal{L}^{\rm sb}_{\rm int}$ \cite{I.15},
\begin{eqnarray}
\label{2.11}
\mathcal{L}_{\rm int} = 
-\left\{ \frac{G_1^{\prime}}{2N}
\left[\left(\bar{\psi}\psi\right)^2+\left(\bar{\psi}\gamma_{45}\psi\right)^2 \right]
+\frac{G_2^{\prime}}{2N} \left[
\left(\bar{\psi}i\gamma_5\psi\right)^2 +
\left(\bar{\psi}i\gamma_4\psi\right)^2 \right] \right\} ~, 
\end{eqnarray}
where
\begin{eqnarray}
 G^{\prime}_1=\frac{1}{4}\left(G^C-\widetilde{G}\right), ~ G^{\prime}_2=\frac{1}{4}\left(G^C+\widetilde{G}\right)~.
 \label{2.12}
\end{eqnarray}

Comparing this with the general expression (\ref{2.3}), one has
$G_1 = G_2 = G^\prime_1,  H_1 = H_2 = G^\prime_2$.  
Neglecting further the symmetry-breaking term (\ref{2.10}), 
i.e. setting $\widetilde{G} = 0$, one gets the symmetric case 
$G^\prime_1 = G^\prime_2$, i.e. $G_i = H_i$  $(i=1,2)$, 
reflecting the $SU(2)$-symmetry of the electromagnetic
Coulomb interaction.

Clearly, the employed approximation scheme does not allow a
determination of the effective coupling constants $G'_{1,2}$ arising
from an underlying microscopic lattice theory, eventually
including lattice vibrations. Thus, the above four-fermion
interaction (\ref{2.11}) can only be considered as a schematic
one. In the following we shall therefore consider its most general
form (\ref{2.3}) and specify, for illustrations, to more symmetric
cases, if necessary. Before concluding this subsection, let us yet
add a remark on Lorentz-invariance.  Clearly, due to the
suppression of spatial components of currents and related
retardation effects in the considered four-fermion interaction in
graphene, the chiral invariant Coulomb expression (\ref{2.9}) and
its Fierz transformation is not Lorentz-invariant. However, this
does not matter here, since (discarding again resulting
vector/axial-vector interaction terms), the Fierz transformation
of a corresponding Lorentz-invariant Thirring-like vector
interaction reproduces, up to a factor 3, the structure
(\ref{2.11}) $(\widetilde{G} = 0)$ (see, e.g., Eqs. (A.8), (A.9) of  App. A
in Ref.\cite{I.15}).

\section{Grand canonical partition function}
\label{sec:3}

The thermodynamics of the considered generalized GN-model at
finite temperature $T$ and chemical potential $\mu$ is most
conveniently described by the grand canonical partition function
\begin{eqnarray}
\mathcal{ Z}(T,\mu) = {\rm Tr}{\rm e}^{-\beta(H-\mu Q)}. 
 \label{3.1}
\end{eqnarray}
Here $H$ is the Hamiltonian of the system, $Q=\bar{\psi}\gamma_3\psi $ 
is the charge operator  and $\beta =1/T$ with $T$ being the temperature. 
It is well-known that $\mathcal{Z}(T,\mu)$ can be expressed as a path integral,
\begin{eqnarray}
 \mathcal{Z}(T,\mu) = \int \mathcal{D}\bar{\psi}\mathcal{D}\psi \exp\left[-\int d^3x \mathcal{L} \right]~, 
 \label{3.2}
\end{eqnarray}
where $\mathcal{L}$ is the model Lagrangian (\ref{2.1})-(\ref{2.3}), and
\begin{eqnarray}
 \int d^3x = \int_{A} d^2\vec{x} \int_{0}^{\beta}d\tau 
\label{3.3}
\end{eqnarray}
denotes integration over the imaginary time $(\tau= it)$ interval
$\tau \in [0,\beta = 1/T]$   and the plane with area $A = l^2$.
Note that the fermionic Grassmann fields in Eq.~(\ref{3.2}) satisfy
$\it{antiperiodic}$ boundary conditions $\psi(\vec{x},0)
=-\psi\left(\vec{x},\beta\right)$.

As usual, performing the fermionic path integral (\ref{3.2}) with
four-fermion interactions requires a Hubbard-Stratonovich
transformation by introducing composite scalar ($\sigma_1$, $\sigma_2$) 
and pseudoscalar ($\varphi_1$, $\varphi_2$) exciton fields. 
The transformed Lagrangian of the generalized GN-model then takes the
form
\begin{eqnarray}
 \label{3.4}
 \mathcal{L}\left[\bar{\psi}, {\psi}, \sigma_i, \varphi_i\right]
&=& N \sum_{k=1}^2 \left[\frac{\sigma_k^2}{2G_k}+\frac{\varphi_k^2}{2H_k} \right]
\nonumber \\
&&+\bar{\psi}\left[\gamma \partial - \mu \gamma_3 + (m_0+\sigma_1) + \sigma_2 \gamma_{45} 
+ \varphi_1 i \gamma_5 + \varphi_2 i \gamma_4\right] {\psi}~.
%\nonumber\\
\end{eqnarray}
Obviously,  by inserting the field equations for exciton fields $(i = 1,2)$
\begin{eqnarray}
 \label{3.5}
\sigma_i &=& - \frac{G_i}{N} \bar{\psi} \Gamma_i \psi~,~\Gamma_i=\left\{I,\gamma_{45} \right\}~,\nonumber\\
\varphi_i &=& - \frac{H_i}{N} \bar{\psi} \widetilde{\Gamma}_i \psi~,~\widetilde{\Gamma}_i=\left\{i\gamma_{5},i\gamma_{4} \right\}~,
\end{eqnarray}
back into expression (\ref{3.4}) reproduces the original form of
the Lagrangian $\mathcal{L}$ in Eq.~(\ref{2.1}) 
\footnote{Note that the factor $1/4$ at the Gaussian term of exciton fields, appearing in Eq.~(40) of Ref.~\cite{I.15}
is incorrect and has to be replaced by $1/2$, as shown in Eq.~(\ref{3.4}) of this text. Correspondingly, Eq.~(\ref{3.5}) 
does not contain a factor 2.}. 
Note that for a constant field $\sigma_1\equiv\langle\sigma_1\rangle$, 
$m = m_0 + \langle\sigma_1\rangle$ can obviously be interpreted as a
dynamical fermion mass.

It is worth remarking that taking the grand canonical expectation
values (condensates $\langle \dots \rangle$) of both sides of Eq.(\ref{3.5}) 
relates condensates of exciton fields 
%$\langle \sigma_i, \varphi_i \rangle$
($\sigma_i, \varphi_i $)
 to condensates of corresponding fermion bilinears
$\langle\bar{\psi}\Gamma_A\psi\rangle$,
$\Gamma_A=\left(\Gamma_i,\widetilde{\Gamma}_i\right)$.\footnote{Recall that 
$\langle \bar{\psi} \Gamma_A \psi \rangle \colon \hspace{-2mm}= \mathrm{Tr} \left\{\left(\bar{\psi} \Gamma_A \psi \right) 
\exp\left[-\beta(H-\mu Q) \right] \right\}/\mathcal{Z}=\int \mathcal{D}\bar{\psi}\, \mathcal{D}\psi 
\left(\bar{\psi} \Gamma_A \psi \right)\exp\left[-\int d^3x \mathcal{L}\right]/\mathcal{Z}$.
}

Using Eqs.~(\ref{3.2}), (\ref{3.4}) and performing a field shift
$\sigma_1\rightarrow \sigma_1-m_0$, the grand canonical partition
function now takes the form $( \kappa = m_0/G_1)$

\begin{eqnarray}
 \label{3.6}
\mathcal{Z}(T,\mu) &=& \int \prod_{i=1}^2 \mathcal{D}\sigma_i \mathcal{D}\varphi_i 
\exp \left[-\int d^3x N \sum_{k=1}^2 \left(\frac{\sigma_k^2}{2G_k}+\frac{\varphi_k^2}{2H_k} - \delta_{k1}\kappa \sigma_1 \right) \right] \nonumber\\
&&\times \int \mathcal{D}\bar{\psi}\mathcal{D}\psi \exp\left[-\int d^3x\int d^3y \, \bar{\psi}(x)\hat{S}^{-1}(x,y)\psi(y) \right]~,
\end{eqnarray}
where bosonic exciton fields must satisfy {\it periodic} boundary
conditions, $\sigma_k\left(\vec{x},0\right) =\sigma_k\left(\vec{x},\beta\right)$,
$\varphi_k\left(\vec{x},0\right) =\varphi_k\left(\vec{x},\beta\right)$. 
Moreover, the inverse fermion propagator  $\hat{S}^{-1}(x,y)$ in Eq.~(\ref{3.6}) 
is given by
\begin{eqnarray}
 \label{3.7}
\hat{S}^{-1}(x,y)&=& \left(\gamma \partial - \mu \gamma_3 + \sigma_1(x) + \sigma_2 (x)\gamma_{45} 
+ \varphi_1(x) i \gamma_5 + \varphi_2(x) i \gamma_4\right)\delta^{(3)}(x-y) {\bf 1}
\nonumber\\
&=& {S}^{-1}(x,y) {\bf 1}~.
\end{eqnarray}
The resulting Gaussian path integral over fermion fields can
easily be performed and leads to a fermion determinant
\begin{eqnarray}
 \left[\det S^{-1}\right]^N = \exp \left[N {\rm Tr} \ln S^{-1}\right]~,
 \label{3.8}
\end{eqnarray}
where  the Tr-symbol means the functional trace of an operator both in
euclidean coordinates and Dirac space, respectively. The grand
canonical partition function is now expressed in terms of
(bosonic) exciton fields alone and reads
\begin{eqnarray}
 \label{3.9}
 \mathcal{Z}(T,\mu) &=& \int \prod_{i=1}^2 \mathcal{D}\sigma_i \mathcal{D}\varphi_i 
\exp \left\{-N \left[\int d^3x \sum_{k=1}^2 \left(\frac{\sigma_k^2}{2G_k}+\frac{\varphi_k^2}{2H_k} - \delta_{k1}\kappa \sigma_1 \right) 
- {\rm Tr} \ln S^{-1}\right]\right\}~.
\nonumber\\
\end{eqnarray}

\subsection{Mean field approximation}
\label{ssec:3a}

As a first step, we now calculate the determinantal term  
$-{\rm Tr} \ln S^{-1}$ for $\it{constant}$ exciton fields, as considered in the
large $N$ (mean field - "{\rm mf}") saddle point approximation of the
path integral. 
The calculations are conveniently done by
Fourier transforming the fermion fields in the respective path integral
(see e.g. \cite{I.24})
 \begin{eqnarray}
 \label{3.10}
 \psi_\alpha(\vec{x},\tau) = \frac{1}{l}\sum_n \sum_{\vec{p}} {\rm e}^{-i(\omega_n \tau + \vec{p}\cdot\vec{x})} \widetilde{\psi}_{\alpha n}(\vec{p})~,
\end{eqnarray}
where the index $\alpha$ refers here to the (reducible) four-component 
Dirac spinor, $\omega_n=(2n+1)\pi/\beta$, $n = 0, \pm1, \pm2,...$ are
fermionic Matsubara frequencies and $\widetilde{\psi}_{\alpha n}(\vec{p})$ is dimensionless. 
Obviously, in the {\rm mf}- (large $N$ saddle point) approximation  the path integral (\ref{3.9}),
evaluated at constant mf-values $( \sigma_k,\varphi_k)_{\rm mf}$ ,
becomes trivial and the grand canonical partition function
factorizes. The relevant thermodynamic quantity, closely related
to $\mathcal{Z}$, is the thermodynamic potential  per area $A = l^2$  ,
\begin{eqnarray}
 \label{3.11}
 \Omega(T,\mu)= - \frac{1}{\beta l^2} \ln \mathcal{Z}(T,\mu)~.
\end{eqnarray}
Its mean-field expression can be separated into an excitonic
condensate part and a fermionic contribution , respectively,
\begin{eqnarray}
 \label{3.12}
\Omega_{\rm mf}=\Omega_{\rm cond}+\Omega_{\rm f}~,
\end{eqnarray}
where
\begin{eqnarray}
 \label{3.13}
\Omega_{\rm cond}&=&N\sum_{k=1}^2 \left[\left( \frac{\sigma_k^2}{2G_k}+\frac{\varphi_k^2}{2H_k} - \delta_{k1}\kappa \sigma_1\right)_{\rm mf} \right] ~,\\
\Omega_{\rm f}&=& - \frac{1}{\beta l^2} N {\rm Tr} \ln \left( \beta S_{\rm mf}^{-1}\right)~,
 \label{3.14}
\end{eqnarray}
and
\begin{eqnarray}
 \label{3.15}
 S_{\rm mf}^{-1}(\vec{p},\omega_n)=\left(-i\gamma_3 (\omega_n - i\mu) - i \vec{\gamma}\cdot\vec{p} + \sigma_1 + \sigma_2 \gamma_{45} 
+ \varphi_1 i \gamma_5 + \varphi_2 i \gamma_4 \right)_{\rm mf}
\end{eqnarray}
is the inverse mean-field fermion propagator in $(\vec{p},\omega_n)$ space.
Obviously, $\Omega_{\rm mf}= \mathcal{O}(N)$   for large $N$. 
Note that the exciton mean field values 
$(\sigma_k,\varphi_k)_{\rm mf} \equiv \left(\bar{\sigma}_k(T,\mu),\bar{\varphi}_k (T,\mu)\right)$ 
at finite temperature $T$ and chemical potential $\mu$ are obtained
as minima (stationary points) of the thermodynamic potential per flavor species
\begin{eqnarray}
 \label{3.16}
 0&=&\frac{1}{N} \frac{\partial\Omega}{\partial\sigma_k}\Bigg{|}_{\rm mf}
 =\left(\frac{\sigma_k}{G_k} -\kappa \delta_{k1} - \frac{1}{\beta l^2} {\rm Tr} S \frac{\partial S^{-1}}{\partial \sigma_k}\right)_{\rm mf}~,\\
0&=&\frac{1}{N} \frac{\partial\Omega}{\partial\varphi_k}\Bigg{|}_{\rm mf}
 =\left(\frac{\varphi_k}{H_k} - \frac{1}{\beta l^2} {\rm Tr} S \frac{\partial S^{-1}}{\partial \varphi_k}\right)_{\rm mf}~,
 \label{3.17}
\end{eqnarray}
where
$\frac{1}{N}\frac{\partial^2\Omega}{\partial\sigma_k^2}\bigg{|}_{\rm mf}>0$,
$\frac{1}{N}\frac{\partial^2\Omega}{\partial\varphi_k^2}\bigg{|}_{\rm mf}>0$ 
and
\begin{eqnarray}
 \label{3.18}
\frac{\partial S^{-1}}{\partial \sigma_k}\Bigg{|}_{\rm mf}&=& \Gamma_k~,~~k=1,2~,\\ 
\frac{\partial S^{-1}}{\partial \varphi_k}\Bigg{|}_{\rm mf}&=& \widetilde{\Gamma}_k~,~~k=1,2~,
 \label{3.19}
\end{eqnarray}
with $\Gamma_k$, $\widetilde{\Gamma}_k$ being defined in Eq.~(\ref{3.5}). 
From Eq.~(\ref{3.5})  it follows that the mean field values (condensates) 
$(\bar{\sigma}_k,\bar{\varphi}_k ) \equiv \left(\langle \sigma_k\rangle,\langle \varphi_k\rangle\right)$ 
correspond to the fermion condensates
$\langle\bar{\psi}\Gamma_k\psi\rangle$, $\langle\bar{\psi}\widetilde{\Gamma}_k\psi\rangle$.

Let us next consider the fermion contribution (\ref{3.14}).
Performing the functional trace ${\rm Tr}$ which is now defined as a sum
over Matsubara frequencies $\omega_n$, two-momenta $\vec{p}$ and
including the trace ${\rm tr}$ over Dirac indices, i.e.
\begin{eqnarray}
 \label{3.20}
{\rm Tr}= \sum_{n,\vec{p}}{\rm tr} = l^2 \sum_n \int \frac{d^2p}{(2\pi)^2} {\rm tr}~,
\end{eqnarray}
we obtain
\begin{eqnarray}
 \label{3.21}
\Omega_{\rm f}&=& - \frac{N}{\beta} \sum_n \int \frac{d^2p}{(2\pi)^2} {\rm tr} \ln \left( \beta S_{\rm mf}^{-1}\right)
= - \frac{N}{\beta} \sum_n \int \frac{d^2p}{(2\pi)^2} \sum_i \ln \left( \beta \varepsilon_i \right)~.
\end{eqnarray}
Here $\varepsilon_i$ are the four eigenvalues of the $4\times 4$
matrix $S_{\rm mf}^{-1}$ which are given by (see \cite{I.7,I.15})
\begin{eqnarray}
 \label{3.22}
 \varepsilon_{1,2,3,4}=\bar{\sigma}_1\pm \sqrt{\left(\bar{\sigma}_2\pm \sqrt{-(\omega_n-i\mu)^2-\vec{p}^2}\right)^2-\bar{\varphi}_1^2-\bar{\varphi}_2^2}~.
\end{eqnarray}
Then,   $\Omega_{\rm f}$  takes the form
\begin{eqnarray}
 \label{3.23}
 \Omega_{\rm f}&=& - \frac{N}{\beta} \sum_{k=1}^2 \int \frac{d^2p}{(2\pi)^2} \sum_n \ln \left\{\beta^2 \left[ (\omega_n-i\mu)^2+{E_{p}^{(k)}}^2 \right]\right\}~,
\end{eqnarray}
where
\begin{eqnarray}
 E_p^{(k)} = \sqrt{\vec{p}^2+M_k^2} 
 \label{3.24}
\end{eqnarray}
and
\begin{eqnarray}
 \label{3.25}
M_{1,2}=\bar{\sigma}_2 \pm \bar{\varrho}~,~\bar{\varrho}=\sqrt{\bar{\sigma}_1^2+\bar{\varphi}_1^2+\bar{\varphi}_2^2}
\end{eqnarray}
are energies or masses associated with two (correspondingly projected) quasiparticle fields,
$\psi_{1,2}=\frac{1}{2}\left(1\pm\gamma_{45}\right)\psi$.
\\
Using standard techniques for performing the Matsubara  summation
of the logarithmic terms in Eq.(\ref{3.23}) (see e.g. \cite{I.24}), we arrive at the final expression
\begin{eqnarray}
 \label{3.26}
  \Omega_{\rm f}&=& - N \sum_{k=1}^2 \int \frac{d^2p}{(2\pi)^2} \left[E_p^{(k)} + \frac{1}{\beta} \ln \left( 1+ {\rm e}^{-\beta(E_{p}^{(k)}-\mu)} \right)
  + \frac{1}{\beta} \ln \left( 1+ {\rm e}^{-\beta(E_{p}^{(k)}+\mu)} \right)\right]~.
\end{eqnarray}

\subsubsection{Vacuum gap equations}

Obviously, the first term in the square bracket of Eq.(\ref{3.26})
is just the vacuum contribution at $T = 0$, $\mu= 0$, whereas the
other terms describe the contributions of a free gas of two types
of  $N$ quasiparticles (anti-quasiparticles) of masses $M_1$ and
$M_2$ at $T$ and $\mu$. Note that the vacuum contribution contains
an UV-divergent momentum integral which will be regularized by a
cutoff $\Lambda$, using polar coordinates for momenta, i.e. $
\left|\vec{p}\right| < \Lambda$. The regularized vacuum
contribution to the thermodynamic potential
$\Omega_{\rm vac}\left(T=0, \mu=0\right)$,
\begin{eqnarray}
 \label{3.27}
 \Omega_{\rm vac}^{\rm reg}(0,0)= N\sum_{k=1}^2 \left[\left( \frac{\bar{\sigma}_k^2}{2G_k}+\frac{\bar{\varphi}_k^2}{2H_k} - \delta_{k1}\kappa \bar{\sigma}_1\right) - \int_{\Lambda} \frac{d^2p}{(2\pi)^2} E_p^{(k)}\right]~,
\end{eqnarray}
then takes the form (comp. \cite{I.7,I.15})
\begin{eqnarray}
 \label{3.28}
 \Omega_{\rm vac}^{\rm reg}(0,0)= N\sum_{k=1}^2 \left[\frac{1}{2}\left( G_k^{-1} - \frac{\Lambda}{\pi}\right) \bar{\sigma}_k^2
 +\frac{1}{2}\left(H_k^{-1} - \frac{\Lambda}{\pi} \right)\bar{\varphi}_k^2 - \delta_{k1}\kappa \bar{\sigma}_1 +\frac{|M_k|^3}{6\pi}\right]~, \nonumber\\
\end{eqnarray}
where, supposing $M_k \ll \Lambda$,  a term $|M_k|^3 \mathcal{O}\left(\frac{M_k}{\Lambda}\right)$ was omitted.

It is convenient to eliminate the cutoff parameter $\Lambda$ in Eq.~(\ref{3.28}) by introducing renormalized coupling constants 
$G^r_k(\hat{\mu})$, $H^r_k(\hat{\mu})$, defined by corresponding normalization conditions at the normalization point $\hat{\mu}$ $(k=1,2)$
\begin{eqnarray}
 \label{3.29a}
 \frac{1}{N}\frac{\partial^2 V}{\partial \bar{\sigma}^2_k}\bigg|_{\bar{\sigma}_k=\hat{\mu}} &=&  \frac{1}{G_k(\Lambda)} - \frac{\Lambda}{\pi} 
 + \frac{2\hat{\mu}}{\pi} \equiv \frac{1}{G^r_k(\hat{\mu})}~,\\ 
 \frac{1}{N}\frac{\partial^2 V}{\partial \bar{\varphi}^2_k}\bigg|_{\bar{\varphi}_k=\hat{\mu}} &=&  \frac{1}{H_k(\Lambda)} - \frac{\Lambda}{\pi} 
 + \frac{2\hat{\mu}}{\pi} \equiv \frac{1}{H^r_k(\hat{\mu})}~.
\label{3.29b}
\end{eqnarray}
Here $V\equiv  \Omega_{\rm vac}^{\rm reg}(0,0)$ denotes the effective potential, and the second derivative in Eq.~(\ref{3.29a}) has to be 
taken at $\bar{\sigma}_k=\hat{\mu}$, $\bar{\sigma}_j=0~(j\neq k)$, $\bar{\varphi}_k=0$  etc.\\
Using Eqs.~(\ref{3.29a}) and (\ref{3.29b}), one can consider new coupling-type parameters $g_k$, $h_k$, 
\begin{eqnarray}
 \label{3.30a}
 \frac{1}{g_k} &\equiv&  \frac{1}{G_k(\Lambda)} - \frac{\Lambda}{\pi} = \frac{1}{G^r_k(\hat{\mu})} - \frac{2\hat{\mu}}{\pi} ~,\\ 
 \frac{1}{h_k} &\equiv&  \frac{1}{H_k(\Lambda)} - \frac{\Lambda}{\pi} = \frac{1}{H^r_k(\hat{\mu})} - \frac{2\hat{\mu}}{\pi}~.
\label{3.30b}
\end{eqnarray}
Obviously, the bare coupling constants $G_k(\Lambda)$ and  $H_k(\Lambda)$ (considered now as functions of $\Lambda$) are independent of the normalization point $\hat{\mu}$. \\
It thus follows from Eqs.~(\ref{3.29a})-(\ref{3.30b}) that the new coupling-type parameters $g_k$ and $h_k$ are also independent of $\hat{\mu}$
and, using analogous arguments, also independent of $\Lambda$. 
For brevity, the quantities $g_k$ and $h_k$ will be in the following called coupling constants.

It is then possible to rewrite the expression (\ref{3.28}) in the form of an ultraviolet-finite, renormalization-invariant quantity
\begin{eqnarray}
 \label{3.29}
\Omega_{\rm vac}^{\rm ren}(0,0)= N\sum_{k=1}^2 \left[\frac{1}{2 g_k} \bar{\sigma}_k^2
 +\frac{1}{2 h_k}\bar{\varphi}_k^2 - \delta_{k1}\kappa \bar{\sigma}_1 +\frac{|M_k|^3}{6\pi} \right]~.
\end{eqnarray}
Note that one gets attractive forces with $g_k<0$, $h_k<0$, if $G_k(\Lambda)>\pi/\Lambda$, $H_k(\Lambda)>\pi/\Lambda$ or,
expressed in terms of renormalized quantities, for $G^r_k(\hat{\mu})>\pi/(2\hat{\mu})$, $H^r_k(\hat{\mu})>\pi/(2\hat{\mu})$.\\
It is instructive to compare this with results for the renormalization-group $\beta-$functions, expressed in terms of rescaled 
dimensionless coupling constants $(\widetilde{G}^r_k,\widetilde{H}^r_k)\equiv (\hat{\mu}{G}^r_k,\hat{\mu}{H}^r_k)$.\\
Using Eqs.~(\ref{3.30a}), (\ref{3.30b}) and $\partial g_k/ \partial \hat{\mu}=\partial h_k/ \partial \hat{\mu}=0$, one obtains
\begin{eqnarray}
 \label{3.31a}
 \beta(\widetilde{G}^r_k)&=& \hat{\mu}\frac{ \partial }{\partial \hat{\mu}} \widetilde{G}^r_k 
 = \widetilde{G}^r_k \left(1 - \widetilde{G}^r_k/\widetilde{G}^{*r}_k \right)~,\\
  \beta(\widetilde{H}^r_k)&=& \hat{\mu}\frac{ \partial }{\partial \hat{\mu}} \widetilde{H}^r_k 
 = \widetilde{H}^r_k \left(1 - \widetilde{H}^r_k/\widetilde{H}^{*r}_k \right)~,
\label{3.31b}
\end{eqnarray}
where $\widetilde{G}^{*r}_k=\widetilde{H}^{*r}_k=\pi/2$.
\\[5mm]
Eqs.~(\ref{3.31a}), (\ref{3.31b}) generalize the results for the usual GN-model (see first paper of Ref.~\cite{I.9}) to the many-coupling case.
In particular, we find that the quantities ${G}^{*r}_k={H}^{*r}_k=\pi/(2\hat{\mu})$ correspond to nontrivial UV fixed points of the generalized, 
graphene-motivated GN-model considered here.

Now, having demonstrated that the expression $\Omega_{\rm vac}^{\rm ren}(0,0)$ is UV-finite,
it is clear that the condensate values
$\bar{\sigma}_k(0,0), \bar{\varphi}_k(0,0)$ at $T = \mu = 0$,
following from the gap equations (\ref{3.16}), (\ref{3.17}), are
described in terms of finite parameters $g_k$, $h_k$.
Due to the explicit symmetry-breaking mass-like term $\kappa$
in Eq.~(\ref{3.29}), the condensate $\bar{\sigma}_1$ depends not
only on $g_1$, but also on $\kappa$. 
\\
Finally, it is worth to remark that in condensed-matter systems like graphene, the cutoff parameter $ \Lambda$ 
is an effective quantity, restricting the low-energy region of applicability of the considered model. 
In particular, $\Lambda = \mathcal{O}(1/a)$ is of the order of the inverse lattice spacing $a$.
%%%%%%%%%%%%%%%%%%%%%%%%%

For illustrations, let us now consider the interesting case of an
effective Coulomb interaction alone (compare discussions below
Eq.~(\ref{2.11}), where now $G^\prime_1 = G^\prime_2 = 1/4 G^C \equiv G$ so
that $g_k = h_k = g$. Moreover, supposing $G > {\pi}/{\Lambda}$ , the renormalized coupling is attractive, 
$g< 0$. As shown in Appendix B, in this case we get $\bar{\sigma}_2=\bar{\varphi}_k=0$. 
The nonvanishing condensate $\bar{\sigma}_1$ satisfies the gap equation (\ref{B5}) and is given by
\begin{eqnarray}
 \label{3.31}
 m\equiv \bar{\sigma}_1 = - \frac{\pi}{2g} \stackrel{+}{(-)}\sqrt{\left(\frac{\pi}{2g} \right)^2+\pi \kappa}
 = M + \kappa |g| + \mathcal{O}(\kappa^2)~,
\end{eqnarray}
where we introduced the mass-like term  $M ={\pi}/{|g|}$. Note
that the chosen positive solution realizes the total minimum of $\Omega_{\rm vac}$.

\subsubsection{Stationarity (gap) equations for $T \neq 0$,  $\mu \neq 0$ }

Using Eqs.~(\ref{3.12}), (\ref{3.13}), (\ref{3.26}), (\ref{3.29}),
one obtains the renormalized version of the thermodynamic
potential
\begin{eqnarray}
 \label{3.32}
 \Omega^{\rm ren}(T,\mu)=\Omega_{\rm vac}^{\rm ren} 
 - \frac{N}{\beta} \sum_{k=1}^2 \int \frac{d^2p}{(2\pi)^2} \left[ \ln \left( 1+ {\rm e}^{-\beta(E_{p}^{(k)}-\mu)} \right)
  + \ln \left( 1+ {\rm e}^{-\beta(E_{p}^{(k)}+\mu)} \right)\right]~,
\end{eqnarray}
where  the couplings $g_k$, $h_k$ appearing in
$\Omega_{\rm vac}^{\rm ren}$ are defined in Eqs.~(\ref{3.30a}), (\ref{3.30b}). 
From  Eqs.~(\ref{3.16}), (\ref{3.17}) and (\ref{3.32}) we obtain the
following general expressions of stationarity (gap) equations ($\eta_k = \pm 1$ for $k=1,2$)
\begin{eqnarray}
 \label{3.33}
&& \frac{\bar{\sigma}_1}{\bar{\varrho}}\left\{\frac{\bar{\varrho}}{g_1} +  \frac{(\bar{\sigma}_2+\bar{\varrho})^2}{2\pi} 
- {\rm sign} (\bar{\sigma}_2-\bar{\varrho}) \frac{(\bar{\sigma}_2-\bar{\varrho})^2}{2\pi} 
+ \right.
\nonumber\\
&& 
\left. \hspace{4cm}
+ \sum_{k=1}^2 \int \frac{d^2p}{(2\pi)^2} \eta_k \frac{M_k}{E_p^{(k)}} \left[f^+(E_p^{(k)})+f^-(E_p^{(k)}) \right]
 \right\} = \kappa~, 
\\
 \label{3.34}
&&\frac{\bar{\varphi}_i}{\bar{\varrho}}\left\{\frac{\bar{\varrho}}{h_i}  +  \frac{(\bar{\sigma}_2+\bar{\varrho})^2}{2\pi} 
- {\rm sign} (\bar{\sigma}_2-\bar{\varrho}) \frac{(\bar{\sigma}_2-\bar{\varrho})^2}{2\pi}
+ \right. \nonumber\\ 
&& \left. \hspace{4cm}
+\sum_{k=1}^2 \int \frac{d^2p}{(2\pi)^2} \eta_k \frac{M_k}{E_p^{(k)}} \left[f^+(E_p^{(k)})+f^-(E_p^{(k)}) \right]
 \right\}=0~,
 \\
&&\left\{\frac{\bar{\sigma}_2}{g_2} +  \frac{(\bar{\sigma}_2+\bar{\varrho})^2}{2\pi} 
- {\rm sign} (\bar{\sigma}_2-\bar{\varrho}) \frac{(\bar{\sigma}_2-\bar{\varrho})^2}{2\pi}
+ \right. 
 \nonumber\\
&& \left. \hspace{4cm}
+\sum_{k=1}^2 \int \frac{d^2p}{(2\pi)^2} \frac{M_k}{E_p^{(k)}} \left[f^+(E_p^{(k)})+f^-(E_p^{(k)}) \right]
 \right\}=0~. 
 \label{3.35}
\end{eqnarray}
Here   $f^{\pm}(E)$ are Fermi-Dirac
distribution functions,
\begin{eqnarray}
f^\pm(E)= \frac{1}{{\rm e}^{\beta (E\pm \mu)}+1}~,
\label{3.36}
\end{eqnarray}
and ${\rm sign}(x)$ is the sign function.

For illustrations, let us again consider the particular case of
symmetric coupling constants $g_k = h_k \equiv g$, admitting,
however, an explicit symmetry-breaking term $\kappa=m_0/G$ related to a
finite bare fermion mass $m_0$. From Eqs.~(\ref{3.33}),
(\ref{3.34}) we then get $\bar{\varphi}_i = 0$,
$\bar{\rho}=\bar{\sigma}_1$ in analogy to the vacuum case (comp. App. B).
Let us in more detail consider the equations for the condensates $\bar{\sigma}_1$, $\bar{\sigma}_2$.
By adding or subtracting Eqs.~(\ref{3.33}) and (\ref{3.35}), one
has ($ M_{1,2} = \bar{\sigma}_2 \pm \bar{\sigma}_1$; $M ={\pi}/{|g|}$ )
\begin{eqnarray}
 \label{3.37}
 \frac{M_1}{\pi}\left\{-M+M_1 + 2\pi \int \frac{d^2p}{(2\pi)^2} \frac{1}{E_p^{(1)}} \left[f^+(E_p^{(1)})+f^-(E_p^{(1)}) \right]
 \right\}&=&\kappa~, \\
%\end{eqnarray}
%
%\begin{eqnarray}
 \label{3.38}
\frac{(-M_2)}{\pi}\left\{-M+(-M_2) + 2\pi \int \frac{d^2p}{(2\pi)^2} \frac{1}{E_p^{(2)}} \left[f^+(E_p^{(2)})+f^-(E_p^{(2)}) \right]
 \right\}&=&\kappa~,~(\bar{\sigma}_2<\bar{\sigma}_1)~. 
 \nonumber \\
 \end{eqnarray}
Comparing Eq.~(\ref{3.38}) with (\ref{3.37}), one finds the solution
$-M_2 = \bar{\sigma}_1-\bar{\sigma}_2 = M_1 =
\bar{\sigma}_1+\bar{\sigma}_2$, $E^{(1)}_ p = E^{(2)}_ p$,  and
thus $\bar{\sigma}_2= 0$, $M_{1,2} = \pm \bar{\sigma}_1$.\\
Finally, performing the momentum integral in Eq.~(\ref{3.37}) gives further
\begin{eqnarray}
 \label{3.39}
  \frac{\bar{\sigma}_1}{\pi}\left\{-M+\bar{\sigma}_1 
  + \frac{1}{\beta} \ln \left[1+ 2 {\rm e}^{- \beta \bar{\sigma}_1} \cosh (\beta \mu)+{\rm e}^{- 2 \beta \bar{\sigma}_1} \right]\right\}=\kappa~.
\end{eqnarray}

\begin{figure}[htb]
	\centering\includegraphics[width=.9\textwidth]{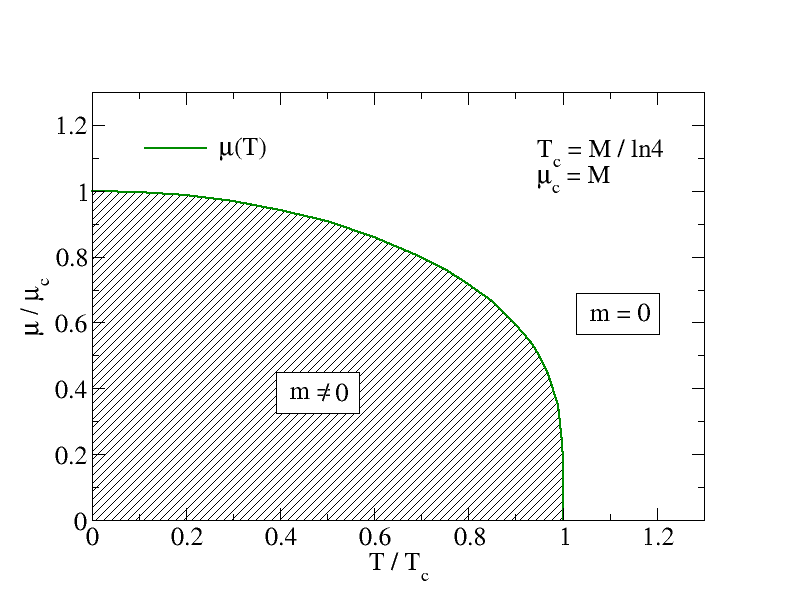}
%	\centering\includegraphics[width=.9\textwidth]{Fig1.png}
	\caption{Phase portrait for the GN-model for $g<0$, $\kappa=0$
		in the $\mu-T$ - plane. The phases of no dynamical symmetry breaking 
		($m=0$) and dynamical symmetry breaking ($m\neq 0$) are separated 
		by a critical line $\mu	(T)$ from Eqs.~(\ref{3.41}), (\ref{3.42}).
		\label{fig:1}}
\end{figure}

Note that in the case of a small bare fermion mass $m_0$ (small
$\kappa$), the system is characterized by a critical curve
corresponding to a phase with a bare fermion mass and a phase with
a larger dynamical mass $m = \bar{\sigma}_1$. In particular, by
setting $\bar{\sigma}_1= 0$, one obtains for the {\it chiral limit} $\kappa = 0$ from Eq.~(\ref{3.39})  an analytical expression for
the critical line of a second-order phase transition, as sketched in the phase portrait of Fig.~\ref{fig:1} (comp. Ref.~\cite{I.8}),

\begin{eqnarray}
-M+\frac{1}{\beta} \ln\left[2+2\cosh(\beta\mu)\right]= 0.
\label{3.40}
\end{eqnarray}

As shown in Ref.~\cite{I.7}, on the critical curve, the
temperature and chemical potential are related by
\begin{eqnarray}
 \label{3.41}
\mu(T)&=&\frac{1}{\beta} \ln K(T)~, \\
% \end{eqnarray}
%
%\begin{eqnarray}
K(T) &=& \left[-1+ \frac{1}{2} \exp(\beta M)\right] + \sqrt{\left[-1+ \frac{1}{2} \exp(\beta M)\right]^2-1}~,    
\label{3.42}
\end{eqnarray}
where  in the chiral limit $M = {\pi}/{|g|}$ is the dynamical
fermion mass $m$ for the vacuum $T = \mu = 0$ (comp. Eq.~(\ref{3.31})). 
Obviously, the function $\mu(T)$ in Eq.~(\ref{3.41}) vanishes for $K(T) = 1$ defining the critical
temperature $T_c$,
\begin{eqnarray}
-1+\frac{1}{2}\exp\left(\beta_c M\right)=1, ~ T_c = M/(2 \ln 2)~.
\label{3.43}
\end{eqnarray}

\subsection{Exciton fluctuations  around mean fields}
\label{ssec:3b}

Up to now we have treated the grand canonical partition function
(\ref{3.9}) in the mean-field (large N saddle point)
approximation. Let us now consider exciton fields as fields
fluctuating around their mean-field values, restricting us to the
case 
$\bar{\sigma}_1 \equiv \langle \sigma_1 \rangle_{\rm mf}\neq0$,
$\bar{\sigma}_2=\bar{\varphi}_1=\bar{\varphi}_2=0$
\footnote{In Refs.~\cite{I.10} there were considered additional excitonic corrections to the gap expression.
Such corrections were, however, shown to contribute in higher order of $1/N$ and will thus be omitted here.
The calculation of $\mathcal{O}(1/N)$ corrections from the composite excitons to the dynamical fermion mass
and of  $\mathcal{O}(1)$ corrections to the fermion polarization function can be straightforwardly done by using 
the variational path integral approach of Ref.~\cite{I.25}.}. 
Then we have
\begin{eqnarray}
 \label{4.1}
 {\sigma}_1(x)&\longrightarrow& \bar{\sigma}_1 + {\sigma}_1(x)~,~~
{\sigma}_2(x)\longrightarrow {\sigma}_2(x)~,~~
{\varphi}_{1,2}(x)\longrightarrow {\varphi}_{1,2}(x)~.
\end{eqnarray}
We can then decompose the inverse fermion propagator $S^{-1}(x,y)$ in Eq. (\ref{3.7})) as
\begin{eqnarray}
% \label{4.2}
 \label{62}
S^{-1}(x,y)&=&(\gamma \partial - \mu \gamma_3 + m) \delta^{(3)}(x-y) + \Sigma (x) \delta^{(3)}(x-y)
\nonumber\\
&=&S^{-1}_{\rm mf}(x-y) + \Sigma (x) \delta^{(3)}(x-y) \equiv S^{-1}_{\rm mf}\left[ 1 + S_{\rm mf}\Sigma \right]~,\\
 \Sigma (x) &=&\sigma_1(x) + \sigma_2(x) \gamma_{45} + \varphi_1(x) i \gamma_5 + \varphi_2(x) i \gamma_4~,
 \label{4.3}
\end{eqnarray}
where $m = \bar{\sigma}_1$. 
In the next step we will perform the path integration in Eq.~(\ref{3.9})  over fluctuating exciton fields in the Gaussian
approximation (assuming again a suitable regularization of the vacuum term). 
For this aim, one has to expand the logarithm in the exponent up to the second (Gaussian) order in the fields
\begin{eqnarray}
 \label{4.6}
-N \left[\int d^3 x \left(\frac{2\sigma_1\bar{\sigma}_1+\sigma_1^2}{2G_1} -\kappa \sigma_1 + \frac{\sigma_2^2}{2 G_2}+ \frac{\varphi_1^2}{2 H_1}+ \frac{\varphi_2^2}{2 H_2}\right) - {\rm Tr} S_{\rm mf}\Sigma + \frac{1}{2}{\rm Tr}\left(S_{\rm mf}\Sigma S_{\rm mf}\Sigma\right) 
\right]~.
\nonumber\\
\end{eqnarray}
Obviously, the linear terms in the fluctuating fields in Eq.~(\ref{4.6}) vanish either due to the gap equation (\ref{3.16})
or due to the corresponding vanishing of  Tr $S_{\rm mf}[ {\Gamma}_i]$ and Tr $S_{\rm mf}[ \widetilde{\Gamma}_i]$. 
Performing then the field integration in the (bosonic) Gaussian integral
\begin{eqnarray}
\mathcal{Z}_{\rm fl} (T,\mu) = \int \prod_{i=1}^2 \mathcal{D}\sigma_i \mathcal{D}\varphi_i 
\exp \left\{- N \left[ \int d^3x \sum_{k=1}^2 \left(\frac{\sigma_k^2}{2G_k}+\frac{\varphi_k^2}{2H_k} \right) 
+ \frac{1}{2}{\rm Tr}\left(S_{\rm mf}\Sigma S_{\rm mf}\Sigma\right)\right] \right\} \nonumber \\
\label{4.7}
\end{eqnarray}
leads to the following factorizing contribution from exciton correlations corresponding to $\sigma_i$ and $\varphi_i$
channels of interactions,
\begin{eqnarray}
\mathcal{Z}_{\rm fl} (T,\mu) = \prod_{i=1,2} \left[ \det \left(\beta \mathcal{D}_{\sigma_i}^{-1}\right) \right]^{-1/2} \cdot
\left[ \det \left(\beta \mathcal{D}_{\varphi_i}^{-1}\right) \right]^{-1/2}~.
\label{4.8}
\end{eqnarray}
Here
\begin{eqnarray}
 \label{4.9}
 \mathcal{D}_{\sigma_i}^{-1}= \frac{N}{G_i} - \Pi_{\sigma_i}~,~~ \mathcal{D}_{\varphi_i}^{-1}= \frac{N}{H_i} - \Pi_{\varphi_i}
\end{eqnarray}
are the inverse (bosonic) exciton propagators containing the
polarization function $\Pi_E$ of the exciton fields $E= (\sigma_i,\varphi_i )$, 
given by corresponding fermion loops, and suitable coordinate dependence is understood.
The calculation of Eq.~(\ref{4.9}) is conveniently done by using Fourier transformation to momentum-frequency space
$p=(\vec{p},\omega_n)$ and $q=(\vec{q},\nu_m)$ for the mean-field propagator $S_{\rm mf} (\vec{p},i\omega_n)$ and the polarization function
 $\Pi_E(\vec{q},i\nu_m)$, where  $\omega_n=(2n+1)\pi / \beta$ and
 $\nu_m=2\pi m / \beta$ $(\{n,m\} = 0, \pm 1, \pm 2,...)$ are the corresponding fermionic and bosonic Matsubara frequencies.
Here the mean-field fermion  propagator reads
\begin{eqnarray}
 S_{\rm mf} (\vec{p},i\omega_n)= \left(-\gamma_3 (i\omega_n + \mu) - i \vec{\gamma}\cdot\vec{p} + m \right)^{-1} 
 = \frac{\gamma_3 (i\omega_n + \mu) + i \vec{\gamma}\cdot\vec{p} + m}{-(i\omega_n + \mu)^2 + E_p^2}~,  
 \label{4.4}
\end{eqnarray}
where
\begin{eqnarray}
E_p = \sqrt{\vec{p}^2+m^2}, 
\label{4.5}
\end{eqnarray}
and we found it now convenient to express the frequency dependence of the propagator by  $i\omega_n$.
\begin{figure}[htb]
	\centering\includegraphics[width=.6\textwidth]{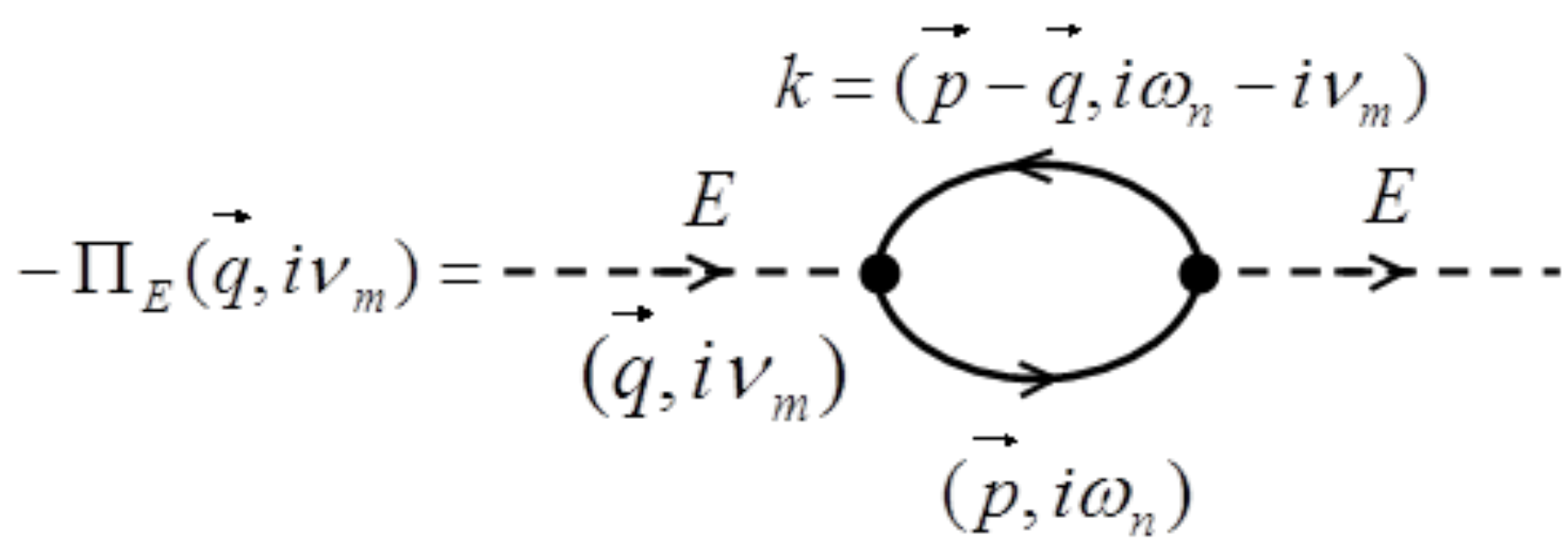}
%	\centering\includegraphics[width=.6\textwidth]{Fig2}
	\caption{Feynman diagram of the irreducible polarization function 
		$\Pi_E(\vec{q},i\nu_m)$ for excitons $E=(\sigma_i,\varphi_i)$.
		The dots describe vertices $\Gamma_E$ given by the Dirac matrices in 
		Eq.~(\ref{4.3}), and the bold lines denote the mean-field fermion propagator $S_{\rm mf} (\vec{p},i\omega_n)$
		of Eq.~(\ref{4.4}).
		\label{fig:2}}
\end{figure}

A straightforward, but lengthy calculation applying Matsubara
summation techniques  to loop diagrams of Fig.~\ref{fig:2}
(see App. C for details) leads to the following expression for the polarization function 
$\Pi_E(\vec{q},i\nu_m)$
\begin{eqnarray}
\Pi_E(\vec{q},i\nu_m)&=& - N\sum_{s,s'=\pm 1}\int {\frac{d^2p}{(2\pi)^2}\frac{f^-(s'E_k)-f^-(s
E_p)}{i\nu_m+s'E_k-sE_p} {\mathcal{T}}^{\mp}}~,
\nonumber\\
 \mathcal{T}^{\mp} &=&
%\left(
1-ss'\frac{\vec{p}\cdot \vec{k}\mp m^2}{E_p E_k}~.
%\right)
\label{4.10}
\end{eqnarray}
Here the minus sign in $\mathcal{T}^{\mp}$ refers to the $\sigma_{1,2}$
excitons, whereas the plus sign refers to $\varphi_{1,2}$ and
$f^{\pm}(E)$ are the Fermi-Dirac distribution functions introduced
in Eq.~(\ref{3.36}).
Obviously, the thermodynamic potential for exciton fluctuations
follows then from 
%Eq.~(\ref{4.8}) 
the Fourier-transformed expression (\ref{4.8}) to be given by
\begin{eqnarray}
 \label{4.11}
 \Omega_{\rm fl} (T,\mu) = \frac{1}{2}\frac{1}{\beta l^2} {\rm \widetilde{Tr}} 
 \left\{\sum_{i=1,2} \left[ \ln \left(\beta \mathcal{D}_{\sigma_i}^{-1}(\vec{q},i\nu_m)\right) 
+ \ln \left(\beta \mathcal{D}_{\varphi_i}^{-1}(\vec{q},i\nu_m)\right)\right] \right\}~,
\end{eqnarray}
where the ${\rm \widetilde{Tr}}$ - symbol refers to summation in $(\vec{q},\nu_m)$ space.
Consider now the total thermodynamic potential per flavor species
\begin{eqnarray}
 \label{4.12}
\frac{1}{N}\Omega(T,\mu) =\frac{1}{N} \left(\Omega_{\rm mf} + \Omega_{\rm fl} \right)~.
\end{eqnarray}
Recalling that $\Omega_{\rm mf} = \mathcal{O}(N)$, the mean-field contribution
in Eq.~(\ref{4.12}) is $\mathcal{O}(1)$, whereas the exciton fluctuating part $\Omega_{\rm fl}$
contributes in the large $N$ limit to (relative) order $\mathcal{O}(1/N)$,
apart from an additive term $\mathcal{O}(\ln N/N)$ independent of $m$.

\section{Equations for bound and resonant exciton states}
\label{sec:4}

\subsection{Bound states}
\label{ssec:4a}

For studying the corresponding equations for excitons in planar
systems, we will closely follow here the analogous techniques of
Refs.~\cite{I.17}, \cite{I.18}, \cite{I.21}, used for studying
$(3+1)$-dimensional models of colored quarks. 
The masses $M_E$ of exciton bound states are determined by the poles of the exciton
propagators, or equivalently, from the vanishing of the inverse propagators of excitons 
taken at rest, $\vec{q} = 0$,
\begin{eqnarray}
 \label{5.1}
 \frac{N}{G_E} - \Pi_E(\vec{0},i\nu_m=M_E)=0~,
\end{eqnarray}
where $G_E$ are generic coupling constants ($G_E = \{G_{1,2}; H_{1,2} \}$), 
and the polarization function $\Pi_E$ is analytically continued to real frequencies (energies) $\omega$,
$i\nu_m\rightarrow \omega=M_E+i\eta$. 
For $\vec{q} = 0$ one has $\vec{k} = \vec{p}$, and the only nonvanishing contribution in
Eq.~(\ref{4.10}) emerges for $s \neq s'$ so that we can put $s' =-s$ and $ss' = -1$ in the following. 
The factor $\mathcal{T}^\pm$ then becomes
\begin{eqnarray}
 \label{5.2}
 1-ss'\frac{\vec{p}\vec{k}\mp m^2}{E_p E_k} = 1+\frac{\vec{p}^2\mp m^2}{E_p^2}
 =\left\{
 \begin{array}{ll}
 {2\vec{p}^2}/{E_p^2}&{\rm for}~\Pi_{\sigma_{1,2}}\\
 2&{\rm for}~\Pi_{\varphi_{1,2}}
 \end{array}
 \right.~,
\end{eqnarray}
and the polarization function is equal to
\begin{eqnarray}
 \label{5.3}
 \Pi_E(\vec{0},M_E)&=& - 2N\sum_{s}\int \frac{d^2p}{(2\pi)^2}\frac{f^-(-sE_p)-f^-(sE_p)}{M_E-s2E_p}
 \left( 
 \begin{array}{c}
 {{p}^2}/{E_p^2}\\
 1
 \end{array}
 \right)~.
\end{eqnarray}
Using the relation
\begin{eqnarray}
f^{\pm}\left(-E_p\right)=1-f^{\mp}\left(E_p\right)~, 
\label{5.4}
\end{eqnarray}
performing the summation over the sign variable $s$ and inserting
the resulting expression into Eq.~(\ref{5.1}), one finally arrives
at the (regularized) bound state equation
\begin{eqnarray}
 \label{5.5}
 \frac{N}{G_E}+2N\int_{\Lambda} \frac{d^2p}{(2\pi)^2}\left[1-f^+(E_p)-f^-(E_p)\right]\left(\frac{1}{M_E-2E_p}-\frac{1}{M_E+2E_p}\right)
 \left( 
 \begin{array}{c}
 {{p}^2}/{E_p^2}\\
 1
 \end{array}
 \right)=0~.
 \nonumber\\
\end{eqnarray}
This equation determines the exciton masses $M_E$ as functions of
the coupling strength $G_E$, cutoff $\Lambda$, temperature $T$,
chemical potential $\mu$ and dynamical fermion mass $m(T,\mu)$. In
general, it has to be solved numerically. However, it is
interesting to present its solution also analytically in a closed
form by using, for convenience, regularized integrals and bare
coupling constants and postponing more detailed renormalization
to a subsequent work devoted to the numerical analysis.
\\[5mm]
Let us start with the gap equation (\ref{3.16}) which for the case
$\bar{\sigma}_2= \bar{\varphi}_i=0$ takes the form
\begin{eqnarray}
 \label{5.6}
m(T,\mu) = \kappa G_1+2G_1 \int_{\Lambda} \frac{d^2p}{(2\pi)^2}\frac{m(T,\mu)}{E_p}\left[1-f^+(E_p)-f^-(E_p)\right]~.
\end{eqnarray}
Supposing a small bare fermion mass $m_0$ (small $\kappa$) and thus
neglecting terms $\mathcal{O}(m^2_0)$, a straightforward calculation leads
to the following mass spectra for the parity doublets
$(\sigma_1,\varphi_1)$, $(\sigma_2,\varphi_2)$,
\begin{eqnarray}
 \label{5.7}
 M^2_{\sigma_1}(T,\mu)&=&4m^2(T,\mu) + \frac{\kappa}{m(T,\mu)} N g^2_{\sigma_1}(T,\mu)~,\\
 \label{5.8}
M^2_{\varphi_1}(T,\mu)&=&\left[\frac{\kappa}{m(T,\mu)} + \left(\frac{1}{H_1}-\frac{1}{G_1} \right)\right]N g^2_{\varphi_1}(T,\mu)~,\\
 \label{5.9}
M^2_{\sigma_2}(T,\mu)&=&4m^2(T,\mu) +\left[\frac{\kappa}{m(T,\mu)} + \left(\frac{1}{G_2}-\frac{1}{G_1} \right)\right]N g^2_{\sigma_2}(T,\mu)~,\\
M^2_{\varphi_2}(T,\mu)&=&\left[\frac{\kappa}{m(T,\mu)} + \left(\frac{1}{H_2}-\frac{1}{G_1} \right)\right]N g^2_{\varphi_2}(T,\mu)~.
 \label{5.10}
\end{eqnarray}
Here  $g^2_E (T,\mu) = \left({\partial\Pi_E}/{\partial q^2}|_{q^2=M^2_E}\right)^{-1}$ is the square of the induced
fermion-exciton coupling constant appearing in the numerator of the exciton propagator in the pole approximation 
\footnote{The expression for the exciton propagator contains the squared coupling constant $g^2_E$ and presents more precisely the 
$T-$matrix expression instead of a pure propagator. Nevertheless, for brevity, it is called propagator here since both objects share the essential 
nonperturbative analytic property of the pole and branch cut structure in the complex plane.}
(see, e.g., Eq.~(30) of Ref.~\cite{I.25} for the case $T = \mu = 0$);
\begin{eqnarray}
 \label{5.11}
 g^2_E(T,\mu)=\left\{ 2N\int_{\Lambda} \frac{d^2p}{(2\pi)^2}\frac{\left[1-f^+(E_p)-f^-(E_p)\right]}{E_p\left(4E_p^2-M_E^2\right)}\right\}^{-1}~.
\end{eqnarray}
Note that  $g^2_E$ behaves for large $N$ as  $g^2_E = \mathcal{O}(1/N)$ so
that the exciton masses in Eqs.~(\ref{5.7}) - (\ref{5.10}) remain finite.

As expected, in the symmetric case of Coulomb-induced four-fermion
interactions with $G_1 = G_2 =H_1 = H_2$, the masses of the
$(\sigma_1,\sigma_2)$ and $(\varphi_1,\varphi_2)$ excitons are degenerate. 
In this case we have approximate continuous chiral symmetries of $\gamma_5, \gamma_4$ 
transformations, explicitly broken by the bare fermion mass term $m_0\bar{\psi}\psi$. 
The $\varphi_1,\varphi_2$ excitons are then the would-be Goldstone
bosons of the explicitly broken chiral symmetry.

As discussed in Sect.~\ref{sec:2}, there exists also the alternative case of
a direct symmetry breaking in coupling constants. 
Indeed, the inclusion of a small on-site repulsive interaction term, arising
in the case of a graphene lattice, as well as of phonon-mediated
interactions, leads to an additional symmetry-breaking interaction
which altogether  gives a modification of effective coupling
constants with $G_1 = G_2 = G^\prime_1$, $H_1 = H_2 =G^\prime_2$, but $G_i
\neq H_i$ (see Eqs.~(\ref{2.11}), (\ref{2.12})). 
In this case the masses of the doublets $(\sigma_1,\sigma_2)$ and
$(\varphi_1,\varphi_2)$ remain (approximately) degenerate,
but $(\varphi_1,\varphi_2)$ excitons do not longer have typical
masses of order $\mathcal{O}(\kappa)$ expected for standard would-be Goldstone
bosons.

Finally, it is worth mentioning that there exists the important
Mermin-Wagner-Coleman (MWC) no-go theorem \cite{I.19} which for
$(2+1)$-dimensional systems forbids spontaneous symmetry breaking
(SSB)  of a continuous symmetry for the case of finite
temperature. Then, in the case of an exact chiral symmetry with
$\kappa= m_0 G_1 = 0$, this requires to go beyond the leading $\mathcal{O}(1/N)$
exciton contributions (see Eq.~(\ref{4.12})) obtained in the large $N$ expansion. 
In our case, this would require to take into account
phase fluctuations of exciton fields  related to vortex
excitations and the Kosterlitz-Thouless transition  \cite{I.8,I.26}. 
The consideration of corresponding higher order $1/N$
corrections  for the massless case $m_0 = 0$ is, however, outside
the scope of this paper. In this context,  it is worth to
emphasize that in the case of explicit symmetry breaking, as
considered in this paper, there do not appear dangerous infrared
singularities in loop diagrams due to the lack of massless
particles. Therefore, the MWC-theorem does not apply here (comp.
with discussions in Refs.~\cite{I.10}).

\subsection{Resonant exciton states}

Clearly, an exciton state becomes a resonant state, if its mass
exceeds two times the dynamical fermion mass so that it can decay
into its constituent fermion-antifermion pair. In this case we are
seeking a pole of the Fourier-transformed exciton propagators
$\mathcal{D}_{\sigma_i}$, $\mathcal{D}_{\varphi_i}$, 
%quoted in Eq.~(\ref{4.9}) and 
analytically continued into the second sheet of
the complex frequency plane, $i\nu _n \rightarrow
z=M_E-i\Gamma_E/2$.
\\
Thus, we have to solve the equation
\begin{eqnarray}
 \label{5.13}
 &&\frac{N}{G_E} - \Pi_E(\vec{0},M_E-i\Gamma_E/2)=\nonumber\\
 &&= \frac{N}{G_E} - \left\{4 N I_1 + 2 N \left[(M_E-i\Gamma_E/2)^2-\varepsilon^2_E \right] I_2(\vec{0},M_E-i\Gamma_E/2) \right\} =0~,
\end{eqnarray}
where $\varepsilon_E= \{2m;0\}$ for $E =\{\sigma_1(\sigma_2);\varphi_1(\varphi_2)\}$. 
\\
Here the two integrals $I_1$, $I_2$ are defined by
\begin{eqnarray}
 \label{5.14}
I_1&=&\int_{\Lambda} \frac{d^2p}{(2\pi)^2}\frac{1}{2E_p}\left[1-f^-(E_p)-f^+(E_p)\right] ~,\\
I_2(\vec{0},z)&=&\int_{\Lambda} \frac{d^2p}{(2\pi)^2}\frac{1}{4E^2_p}\left[1-f^-(E_p)-f^+(E_p)\right]\left(\frac{1}{2E_p-z}+\frac{1}{2E_p+z}\right)~,
 \label{5.15}
\end{eqnarray}
and  $z = M_E  - i\Gamma_E/2$.      .

Then, by separating Eq.~(\ref{5.13}) into one for the real part and another one for the imaginary part, one obtains
\begin{eqnarray}
 \label{5.16}
M^2_{E_i}-\frac{1}{4}\Gamma^2_{E_i}-\varepsilon^2_{E_i} &=& \frac{\frac{1}{2G_{E_i}}-2I_1}{|I_2(\vec{0},M_{E_i}) |^2} {\rm Re} I_2(\vec{0},M_{E_i})~,\\
M_{E_i}\Gamma^2_{E_i}&=&\frac{\frac{1}{2G_{E_i}}-2I_1}{|I_2(\vec{0},M_{E_i}) |^2} {\rm Im} I_2(\vec{0},M_{E_i})~.
 \label{5.17}
\end{eqnarray}
The calculation of $ {\rm Re} I_2$ and $ {\rm Im} I_2$ appearing in Eqs.~(\ref{5.16}) and (\ref{5.17}) can be done by using the standard formula
\begin{eqnarray}
 \label{5.18}
 \frac{1}{x-x_0\mp i\eta}={\rm P.V.}\frac{1}{x-x_0}\pm i \pi \delta(x-x_0)~,
\end{eqnarray}
where P.V. means {\it principal value}.
\\
After some lengthy but straightforward calculations, analoguous to those in Ref.~\cite{I.30}, one gets
\begin{eqnarray}
 \label{5.19}
{\rm Re} I_2&=&{\rm P.V.} \int_{\Lambda} \frac{d^2p}{(2\pi)^2}\frac{1}{4E^2_p}\left[1-f^-(E_p)-f^+(E_p)\right]
\left(\frac{1}{2E_p-M_E}+\frac{1}{2E_p+M_E}\right)~,\\
{\rm Im} I_2&=&\frac{1}{8M_E}\left[1-f^-(\frac{M_E}{2})-f^+(\frac{M_E}{2})\right]
\Theta(M_E^2-4m^2)\Theta(4(\Lambda^2+m^2)-M_E^2)~,%\nonumber\\
 \label{5.20}
\end{eqnarray}
where the UV-cutoff $\Lambda$ was used in performing the momentum integration and $\Theta(x)$ is the step function.
\\
In deriving Eqs.~(\ref{5.16}) and (\ref{5.17}) it was assumed that the imaginary part $\Gamma_E$ occuring in the argument $z$ of $I_2$  (which is the full width at the half maximum of the Breit-Wigner distribution) is small compared to the mass $M_E$ and may 
be neglected so that the equations for $M_E$ and $\Gamma_E$ decouple (compare with the discussions in Ref.~\cite{I.17}).

In the symmetric case $G_i=H_i$ one expects on general grounds that with increasing temperature the constituent fermion mass $m(T,\mu)$
and thus the two-particle threshold $2m(T,\mu)$ lowers, whereas the mass of the would-be Goldstone boson $M_{\varphi_1}(T,\mu)$ is chirally protected and thus rather temperature independent.
Therefore it is inevitable that a temperature exists where the mass of $\varphi_1$ coincides with that of its constituents and thus the binding energy 
vanishes. 
This temperature is called {\it Mott temperature} $T_{\rm Mott}$ and is defined by the relation 
\begin{eqnarray}
 \label{5.21}
 M_{\varphi_1}(T_{\rm Mott},\mu)= 2 m(T_{\rm Mott},\mu)~.
\end{eqnarray}
The effect of bound state dissolution caused by a lowering of the continuum edge in a hot and dense medium is called {\it Mott effect} due to 
its analogy with the physics of the insulator-metal transition according to N. Mott \cite{I.33,I.34}. 

\section{Beth-Uhlenbeck approach to the fermion-exciton plasma}
\label{sec:5}

Let us now in more detail investigate the thermodynamic potential   $\Omega_{\rm fl}(T,\mu)$  
for exciton fluctuations, given in Eq.~(\ref{4.11}), by rewriting the logarithm of the inverse exciton propagator as an integral
\begin{eqnarray}
 \label{6.1}
 \ln \left\{\beta \left[\frac{N}{G_E} - \Pi_E(\vec{q},i \nu_m) \right]\right\}&=& 
 \int^{1}_\varepsilon d\lambda \frac{d}{d\lambda}\ln \left\{\beta \left[\frac{N}{\lambda G_E} - \Pi_E(\vec{q},i \nu_m) \right]\right\} + C
\nonumber\\
&=&-\int^{1}_\varepsilon \frac{d\lambda}{\lambda^2} \frac{N}{G_E}\frac{1}{ \left[\frac{N}{\lambda G_E} - \Pi_E(\vec{q},i \nu_m) \right]} + C~,
\end{eqnarray}
considered in the limit $\varepsilon \to 0$, and  $G_E = \{G_i, H_i\}$, $E = \{\sigma_i,\varphi_i\}$.
Note that the frequency-independent constant $C=\ln [\beta N/(G_E \varepsilon)]$ compensates the logarithmic divergence of the integral at the lower boundary $\varepsilon$ and reproduces the logarithmic dependence on $N$ and $\beta$ of the l.h.s.\\
Next, it is convenient to rewrite the exciton propagator $\mathcal{D}^\lambda_E$ with coupling constant $\lambda G_E$ as dispersion relation with the spectral function $A^{\lambda}_E  (\vec{q}, \omega)$ 
\cite{I.27}
\begin{eqnarray}
 \label{6.2}
 -\int^{1}_\varepsilon \frac{d\lambda}{\lambda^2} \frac{N}{G_E}\frac{1}{ \left(\frac{N}{\lambda G_E} - \Pi_E(\vec{q},i \nu_m) \right)} 
 -\ln \varepsilon
 = - \int^{1}_\varepsilon \frac{d\lambda}{\lambda^2} \frac{N}{G_E} \int_{-\infty}^\infty \frac{d\omega}{2\pi} \frac{A^{\lambda}_E(\vec{q},\omega)}{\omega - i \nu_m}~.
\end{eqnarray}
Recalling that the spectral function $A^\lambda_E$ is given by the doubled imaginary part  of the analytic continuation of the propagator to the real   
$\omega$-axis at positions  $\omega\pm i \eta$, where it has branch cuts,  one obtains
\begin{eqnarray}
 \label{6.3}
 \int^{1}_\varepsilon \frac{d\lambda}{\lambda^2} \frac{N}{G_E} A^{\lambda}_E(\vec{q},\omega)
 &=&-i \int^{1}_\varepsilon \frac{dg}{\lambda^2} \frac{N}{G_E}\left[\mathcal{D}^{\lambda}_E(\vec{q},\omega+i\eta) - \mathcal{D}^{\lambda}_E(\vec{q},\omega-i\eta)\right] \nonumber \\
 &=&-i \ln \frac{\frac{N}{G_E} - \Pi_E(\vec{q},\omega-i\eta)}{\frac{N}{G_E} - \Pi_E(\vec{q},\omega+i\eta)}
 = -i \ln \frac{\mathcal{D}_E(\vec{q},\omega+i\eta)}{\mathcal{D}_E(\vec{q},\omega-i\eta)}~.
\end{eqnarray}
It is worth remarking that the argument of the logarithm in the obtained expression is the very definition of  the scattering matrix $S_E$ 
in the Jost representation \cite{I.17},
\begin{eqnarray}
 \label{6.4}
 S_E(\vec{q},\omega)=\frac{\mathcal{D}_E(\vec{q},\omega+i\eta)}{\mathcal{D}_E(\vec{q},\omega-i\eta)}~.
 %=\frac{1-G_E/N~ \Pi_E(\vec{q},\omega-i\eta)}{1-G_E/N~ \Pi_E(\vec{q},\omega+i\eta)}~.
\end{eqnarray}
Because above the threshold for elastic scattering  $\left| S_E \right|  = 1$, one has
\begin{eqnarray}
 \label{6.5}
 S_E (\vec{q},\omega) = {\rm e}^{2i\phi_E}~,
\end{eqnarray}
where $\phi_E$ is the phase of  $\mathcal{D}_E (\vec{q},\omega+ i\eta)$. 
Then, using Eqs.~(\ref{6.1})-(\ref{6.5}), one gets
\begin{eqnarray}
 \label{6.6}
 \ln \left(\beta \mathcal{D}^{-1}_E \right)= \int^\infty_{-\infty} \frac{d\omega}{\pi} \frac{1}{ i \nu_m - \omega}\phi_E(\vec{q},\omega) + C^\prime~,
\end{eqnarray}
where $C^\prime=\ln (\beta N/G_E)$.
\\
Finally, by inserting  the expression (\ref{6.6}) into the fluctuating part of the thermodynamic potential (\ref{4.11})
and performing the trace over momenta and bosonic Matsubara frequencies $\nu_m$
(compare Eq.~(\ref{C12}) in App.~C), one obtains
\begin{eqnarray}
 \label{6.7}
\Omega_{\rm fl}(T,\mu)= - \sum_E \int \frac{d^2q}{(2\pi)^2} \int^\infty_{-\infty} \frac{d\omega}{2\pi} g(\omega) \phi_E(\vec{q},\omega)~,
\end{eqnarray}
where $g(\omega)$ denotes the Bose-Einstein distribution function 
\begin{eqnarray}
 \label{6.8}
               g(\omega) = \left[\exp(\beta\omega) - 1\right]^{-1}~,
\end{eqnarray}
and an additive term $\mathcal{O}(\ln N)$ independent of $m$ is discarded (comp. the discussion below Eq.~(\ref{4.12})).
\\
It is convenient to split the integral (\ref{6.7}) in the following way
\begin{eqnarray}
 \label{6.9}
 \Omega_{\rm fl}(T,\mu)&=&  -\sum_E \int \frac{d^2q}{(2\pi)^2} \left[\int_{-\infty}^0 \frac{d\omega}{2\pi} g(\omega) \phi_E(\vec{q},\omega)
 +{\int_0^{\infty}} {\frac{d\omega}{2\pi} g(\omega) \phi_E(\vec{q},\omega)}\right]\nonumber\\
 &=&-\sum_E \int \frac{d^2q}{(2\pi)^2} {\int_0^{\infty}} \frac{d\omega}{2\pi}  \left[g(-\omega) \phi_E(\vec{q},-\omega)
 +g(\omega) \phi_E(\vec{q},\omega)\right]~.
\end{eqnarray}
Next, by using the properties  $\phi_E(\vec{q},-\omega) \equiv -\phi_E(\vec{q},\omega)$ and $g(-\omega) = - (1 + g(\omega))$,
and performing a partial integration, we obtain the final form of the excitonic thermodynamic potential
\begin{eqnarray}
 \label{6.10}
 \Omega_{\rm fl}(T,\mu)=\sum_E \int \frac{d^2q}{(2\pi)^2} {\int_0^{\infty}} \frac{d\omega}{2\pi} 
 \left[\omega +\frac{2}{\beta} \ln \left(1-{\rm e}^{-\beta \omega} \right)\right] \frac{d \phi_E}{d \omega}~, 
\end{eqnarray}
which is the so-called extended Beth-Uhlenbeck equation. The connection with the original Beth-Uhlenbeck expression for the second virial coefficient in terms of the two-body scattering phase shift can be found in Refs.~\cite{I.20}. 
Combining a possible pole term of the S-matrix and the scattering contribution, gives
\begin{eqnarray}
 \label{6.11}
 \frac{d \phi_E(\vec{q},\omega)}{d \omega} \longrightarrow \pi \delta(\omega-E_E) +  \frac{d \phi_E(\vec{q},\omega)}{d \omega}\bigg|_{\rm scatt}~.
\end{eqnarray}
Note that in the pole approximation only the first term in Eq.~(\ref{6.11}) is taken into account with $E_E = \sqrt{\vec{q}^2 + M_E^2}$.
In this case one obtains from Eq.~(\ref{6.10}) the thermodynamic potential for a (quasi)relativistic Bose gas of non-interacting excitons
\begin{eqnarray}
 \label{6.12}
 \Omega^{\rm gas}_{\rm fl}(T,\mu)= \sum_E \int \frac{d^2q}{(2\pi)^2} \left[\frac{1}{2}E_E +\frac{1}{\beta} \ln \left(1-{\rm e}^{-\beta E_E} \right)\right]~.
\end{eqnarray}
Correspondingly, the pressure of a gas of free excitons is given by
\begin{eqnarray}
 \label{6.13}
P_{\rm fl} = - \Omega^{\rm gas}_{\rm fl}(T,\mu)
= - \sum_E \int \frac{d^2q}{(2\pi)^2} \left[\frac{1}{2}E_E +\frac{1}{\beta} \ln \left(1-{\rm e}^{-\beta E_E} \right)\right]~.
\end{eqnarray}
Here the divergent momentum integral of the excitonic vacuum energy has again to be regularized by a cutoff and should be subtracted 
afterwards (see discussion below).

It is generally convenient to perform the variable transformation $\omega  = \sqrt{\vec{q}^2 + s} $ in Eq.~(\ref{6.10}) and to introduce the 
density of states
\begin{eqnarray}
 \label{6.14}
            D _E(s) = \frac{1}{\pi}\frac{d \phi_E(s)}{d s}~.
\end{eqnarray}
This leads to the alternative representation
\begin{eqnarray}
 \label{6.15}
\Omega_{\rm fl}(T,\mu)=  \sum_E \int \frac{d^2q}{(2\pi)^2} \left\{\int_{-\vec{q}^2}^\infty ds 
\left[\frac{1}{2}\sqrt{\vec{q}^2+s} +\frac{1}{\beta} \ln \left(1-{\rm e}^{-\beta \sqrt{\vec{q}^2+s}} \right)\right]D _E(s)\right\}~.
\end{eqnarray}
The total thermodynamic potential per flavor species is then given by Eqs.~(\ref{4.12}), (\ref{6.15}) .       

As is well-known, the equation of state and relevant thermodynamic relations are encoded in  
$\Omega (T, \mu; m(T,\mu))$, which corresponds to the pressure density up to a sign. Obviously, only the pressure density relative to the physical vacuum term       $\Omega^{\rm phys}_{\rm vac}$  at  $T = \mu = 0$
\begin{eqnarray}
 \label{6.16}
 \Omega^{\rm phys}_{\rm vac}=\Omega^{\rm mf}_{\rm vac}+\Omega_{\rm fl}\bigg|_{T=\mu=0}
\end{eqnarray}
can be measured. Here the mean-field vacuum expression  $\Omega^{\rm mf}_{\rm vac}$ is given in Eq.~(\ref{3.27})
and the term   $\Omega_{\rm fl}\bigg|_{T=\mu=0}$ has to be taken from Eq.~(\ref{6.15}). 
Following an analogous procedure as for (3+1)-dimensional  models of colored quarks in Ref.~\cite{I.17}, the above vacuum term should be subtracted to give the physical thermodynamic potential
\begin{eqnarray}
 \label{6.17}
 \widetilde{\Omega} (T, \mu)=\Omega (T, \mu) - \Omega_{\rm vac}^{\rm phys}~,
\end{eqnarray}
which is again denoted by $\Omega (T, \mu)$ . 
Clearly, such a removal of the divergent vacuum energy contributions leads to a vanishing pressure $P(T,  \mu) =  0$ at $T = \mu = 0$.
This procedure is analogous to the "no sea" approximation in relativistic mean-field approaches  as, e.g., used in the Walecka  model \cite{I.18}.

Finally, let us return to the expression given in Eq.~(\ref{6.15}) and consider the first departure beyond the pole approximation used in 
Eq.~(\ref{6.12}). 
Clearly, not too far above the Mott temperature  $T_{\rm Mott}$  it is justified to describe $D_E (s)$ by a Breit-Wigner type function
(see a related proposal for quark matter in Ref.~\cite{I.18})
\begin{eqnarray}
 \label{6.18}
    D^R_E (s) = a_R \frac{M_E \Gamma_E}{(s-M_E^2)^2+(M_E \Gamma_E)^2}~.
\end{eqnarray}
Here $M_E$ is the exciton pole mass, $\Gamma_E$ is the corresponding width and $a_R$ is a normalization factor. 
Below $T_{\rm Mott}$, where the spectral broadening $\Gamma_E(T)$  of the state vanishes, the above expression becomes a delta function,
typical for the spectral contribution of an exciton bound state (see Eq.~(\ref{6.11}).
The exciton phase shift $\phi^R_E$ corresponding to Eq.~(\ref{6.18}) should then be of the form \cite{I.21}
\begin{eqnarray}
 \label{6.19}
 \phi_E^R(s) = \frac{\pi}{\frac{\pi}{2} - \arctan \left(\frac{4m^2 - M_E^2}{M_E \Gamma_E} \right)}
 \left[\arctan \left(\frac{s - M_E^2}{M_E \Gamma_E} \right) -  \arctan \left(\frac{4m^2 - M_E^2}{M_E \Gamma_E} \right) \right]
\end{eqnarray}
and represents, by construction, the resonant properties of excitonic correlations.
As discussed in detail in the case of a (3+1)-dimensional Nambu-Jona-Lasinio quark model \cite{I.17} and applies also here, the pole approximation alone is insufficient to fulfil Levinson's theorem \cite{I.28}. In our case, it reads
\begin{eqnarray}
 \label{6.20}
  \int_{4m^2}^\infty ds  D_E(s)  = 0~.
\end{eqnarray}
Obviously, Eq.(\ref{6.20}) requires that the density function must take both positive as well as negative values in such a way that the integral 
can vanish. 
As discussed for quark models \cite{I.17,I.21,I.22}, this should be the case, if a corresponding background scattering contribution
%phase shift part  
$\phi^{\rm sc}_E$ 
is taken into account leading to the decomposition
\begin{eqnarray}
 \label{6.21}
 \phi_E=\phi^{\rm R}_E + \phi^{\rm sc}_E~.
\end{eqnarray}
We expect an analogous argumentation to apply to the generalized (2+1)-dimensional GN-model  considered here.

%\section{Numerical results (to be done)}

\section{Summary and conclusions}
\label{sec:6}

In this paper we have extended the investigation of the
generalized, graphene-motivated $(2+1)$-dimensional GN-model of
Ref.~\cite{I.15},  performed at $T=0$,  to  the case of finite
temperature and fermion density (chemical potential) and have been
going also beyond the mean-field approximation. The schematic model
with four different types of four-fermion interactions considered
here arises in a natural way by Fierz-transforming an effective
electromagnetic Coulomb interaction onto four types of $f \bar{f}$-channels. 
The model includes further a small symmetry-breaking
repulsive on-site interaction term and a small bare fermion mass
which explicitly breaks chiral symmetry. 
The bare fermion mass term avoids IR-divergencies and the validity of the MWC-no-go-theorem.

Throughout we have used the powerful tool of large-$N$ path integral
techniques to derive the bosonized version of the grand canonical
partition function and the related thermodynamic potential in
terms of  four composite exciton fields $E=\{\sigma_i,\varphi_i\}$
$(i=1,2)$. We then first employ  the large-$N$ saddle point
approximation to the path integral which is based on the
mean-field values $\{\bar{\sigma}_i,\bar{\varphi}_i \}$. They are
determined from the minimum of the thermodynamic potential in the
form of gap equations which are studied in some detail. The
mean-field approximation just leads to the dominant contribution
in the thermodynamic potential and the related equation of state.
On the other hand, as it is well-known from four-quark models of
strong interactions (see e.g. \cite{I.10}, \cite{I.17},
\cite{I.18}), light composite (would-be) Goldstone mesons (pions)
play an important role in the low temperature region of a
quark-meson plasma, where quarks are heavy due to a nonvanishing
quark condensate. At higher temperatures the dynamical quark mass
decreases, the quark threshold lowers and mesons are expected to
become heavier resonances. Thus, in the high temperature region
fermions should dominate thermodynamics. As emphasized in the
literature, it is just this different behavior in the low and high
$T$ regions that makes it necessary to extend the large-$N$ path
integral approach beyond mean field by including contributions of
fluctuating fields at (relative) order $\mathcal{O}(1/N)$ (\cite{I.10},
\cite{I.17}). 
We followed here the analogous line of argumentation for the chosen graphene-motivated GN$_3$ model. 
In particular, we present the thermodynamic potential of the
resulting fermion-exciton plasma at next order in the 1/$N$
expansion. Besides of general expressions for the fermion and
exciton spectra at finite temperature and densities, one of the
main results is the presentation of the exciton contribution to
the thermodynamic potential in a form which extends the results of
Beth and Uhlenbeck as well as Dashen et al. \cite{I.20} to the in-medium
case. The numerical calculation of the thermodynamic potential and
the numerical estimate of the importance of the $\mathcal{O}(1/N)$
corrections within the underlying generalized GN$_3$ model are under way.

\section*{Acknowledgment}
One of the authors (D.E.) thanks K. G. Klimenko and V. Ch. Zhukovsky
for  a long-term fruitful cooperation and many useful discussions.
D.B. is grateful to N. T. Gevorgyan for assistance with preparing the manuscript.
The authors acknowledge support by the DAAD partnership program
between the Humboldt University Berlin and the University of
Wroclaw and the hospitality extended to them at these
Institutions. D.B. was supported by the Polish Narodowe Centrum Nauki 
(NCN) under grant No. UMO-2014/15/B/ST2/03752 
and by the MEPhI Academic Excellence Program under contract No. 02.a03.21.0005.
%RUDN University Program 5-100.

\appendix

\section{Dirac matrices in $D=3$}
\label{app:A}

\subsection{Algebra}

We find it convenient to use the reducible $4\times 4$ chiral (Weyl)
representation of Dirac matrices arising in a natural way in
low-energy effective Lagrangians of hexagonal graphene-like
lattice sheets (see e.g. \cite{I.15} , \cite{I.29} and
references therein). In euclidean notation, where
$\gamma^e_k=-i\gamma^k$,  $k=1,2$, and $\gamma_3^e=\gamma^0$, 
we have hermitean matrices (omitting now the index $"e"$)
\begin{eqnarray}
\label{A1} 
\gamma_k =\left(
\begin{array}{cc}
0&i\tau_k\\
-i\tau_k&0
\end{array}
\right), ~~\gamma_3 =\left(
\begin{array}{cc}
0&I_2\\
I_2&0
\end{array}
\right)~~(k=1,2)~,
\end{eqnarray}
with $\tau_k$, $I_2$ being $2\times2$ Pauli matrices and the $2\times2$ unit matrix, respectively. 
In addition, there exist two "chiral " $4\times4$ matrices which anticommute with all $\gamma_\mu$ 
($\mu = 1,2,3$) and with each other,
\begin{eqnarray}
\label{A2} 
\gamma_4 =\left(
\begin{array}{cc}
0&i\tau_3\\
-i\tau_3&0
\end{array}
\right), ~~
\gamma_5=\gamma_1\gamma_2\gamma_3\gamma_4 
=\left(
\begin{array}{cc}
I_2&0\\
0&-I_2
\end{array}
\right)~,
\end{eqnarray}
as well as their combination
\begin{eqnarray}
\label{A3} 
\gamma_{45} =\frac{1}{2}[i\gamma_4,\gamma_5]
=\left(
\begin{array}{cc}
0&\tau_3\\
\tau_3&0
\end{array}
\right)~.
\end{eqnarray}
Note that $\gamma_{45}$ commutes with all $\gamma_\mu$, but anticommutes with 
$\gamma_4$ and $\gamma_5$. 
The matrices $\gamma_\mu$ ($\mu= 1,...,4$) satisfy the euclideanized Dirac algebra
\begin{eqnarray}
\label{A4} 
\left\{\gamma_\mu,\gamma_\nu\right\}=2\delta_{\mu\nu}I~,
\end{eqnarray}
where $I$  is the $4\times4$ unit matrix.
It is interesting to note that in the "chiral limit" $m_0 = 0$ ($\mu = 0$) the matrices 
(\ref{A2})-(\ref{A3}) are the generators $t_i$ ,
\begin{eqnarray}
\label{A5} 
t_1=\frac{1}{2}\gamma_4, ~~~ t_2=\frac{1}{2}\gamma_5,~~~ t_3=\frac{1}{2}\gamma_{45}~,
\end{eqnarray}
of a symmetry group $SU(2)$ of the free Lagrangian $\mathcal{L}_0$ in Eq.~(\ref{2.2}) with group algebra
\begin{eqnarray}
\label{A6} 
\left[t_i,t_j\right]=i\varepsilon_{ijk}t_k~,
\end{eqnarray}
where $\varepsilon_{ijk}$ is the Levi-Civita symbol.

\subsection{Relations to graphene}
\label{app:graphene}

For possible applications to planar systems in condensed matter
physics, it is worth to compare with corresponding notations used
in the chiral limit ($m_0 = 0$) in graphene (see \cite{I.15}, \cite{I.29}; 
for an excellent review, see also \cite{I.12} and references therein). 
As is well-known, the above $SU(2)$ group describes there an emergent
global continous "pseudospin-valley" symmetry. 
Moreover, one considers chiral spinors 
$\psi_{\pm}=\frac{1}{2}\left(1\pm \gamma_5\right)\psi$ 
which have chirality eigenvalues $\pm 1$,
\begin{eqnarray}
\label{A7} \gamma_5 \psi_\pm=\pm \psi_\pm.
\end{eqnarray}
They correspond to fermion excitations at the two inequivalent
"Dirac points" $\vec{\rm K}_+$, $\vec{\rm K}_- = - \vec{\rm K}_+$ at the corners of the first
Brillouin zone, where band energies vanish, $\varepsilon_\pm(\vec{\rm K}_\pm) = 0$. 
In particular, by
performing a low-momentum expansion  $\vec{k} = \vec{{\rm
K}_+}(\vec{\rm K}_-) + \vec{p}$ around these points, one obtains a
linear dispersion law $\varepsilon_\pm = \pm {\rm v}_F |\vec{p}|$
for massless quasiparticles on the hexagonal graphene lattice,
enabling a quasirelativistic description by a Dirac Lagrangian. 
Definite chirality eigenvalues $\pm 1$ coincide with
the so-called valley index $\eta=\pm 1$, characterizing the Dirac
points $\vec{\rm K}_+$ and $\vec{\rm K}_-$. Besides of two "Dirac point
(valley)" degrees of freedom, graphene electrons possess in
addition two "sublattice" or "pseudospin" d.o.f. . The latter ones
are associated to the two existing triangular sublattices of
carbon atoms in the hexagonal graphene lattice to which an
electron belongs. By combining the corresponding two valley and
pseudospin (sublattice) d.o.f. into a four-spinor, makes it just
possible to define chiral $4\times 4$ matrices (A2) and to introduce
various types of fermion interactions. Finally, these analogies
then allow us to apply nonperturbative methods of relativistic
QFT, extensively used for studying dynamical mass generation and
phase transitions for quark models of strong interactions in
dependence on various types of symmetry breaking scenarios.

\subsection{Discrete Symmetries}
\label{app:symmetries}

Let us next consider a convenient choice of discrete symmetry transformations $\mathcal{P, T, C}$
based on the condensed-matter roots of the graphene-motivated GN$_3$-model considered in the text.
We shall here follow the definitions of Gusynin et al. \cite{I.29} which have to be slightly modified when 
using euclidean notations and Grassmann fields.

Consider (reducible) four-spinor fields of graphene with components referring to two inequivalent Dirac points 
$\vec{\rm K}_+$, $\vec{\rm K}_- = - \vec{\rm K}_+$ and triangular sublattices with A and B sites of a hexagonal honeycomb 
lattice given by\footnote{Note that Fig.~1 of Ref.~\cite{I.29} for the hexagonal graphene lattice and the Brillouin zone 
with corresponding Dirac points $\vec{\rm K}_\pm$ slightly differs from Fig.~1 of Ref.~\cite{I.15}. 
It is worth to remark that the Fourier transformed Grassmann components $\psi^{A}_{K_+}$, ${\psi^{A}_{K_+}}^{\hspace{-5pt}*}$ etc.
correspond to their one-particle annihilation or creation operators $a_{K_+}$, ${a^\dagger_{K_+}}$ etc.}
\begin{equation}
\psi =\left(
\begin{array}{c}
\psi^{A}_{K_+}\\
\psi^{B}_{K_+}\\
\psi^{B}_{K_-}\\
\psi^{A}_{K_-}
\end{array}
\right)~~,~
\bar{\psi} = \psi^\dagger\gamma_3 = \left({\psi^{B}_{K_-}}^{\hspace{-5pt}*},{\psi^{A}_{K_-}}^{\hspace{-5pt}*},{\psi^{A}_{K_+}}^{\hspace{-5pt}*},{\psi^{B}_{K_+}}^{\hspace{-5pt}*} \right)~.
\end{equation}
(Note that the sublattices A, B were exchanged in components at the $\vec{\rm K}_-$ point in order to get a suitable four-spinor notation.)
For shortness, we omitted here the combined flavor index $a$ including the considered number $\tilde{N}$ of hexagonal monolayers and the number 2 of physical spin degrees of freedom, i.e. $a\equiv (i,\sigma)$ with $i=1, \dots, \tilde{N}$ and
$\sigma=(\uparrow, \downarrow)$ so that $a=1,\dots, N=2\tilde{N}$.
Moreover, " * " denotes here the operation of Grassmann involution. 

\subsubsection{Spatial inversion $\mathcal{P}$}

By definition,  $\mathcal{P}$ exchanges points $x=(\vec{x},x_3)\to \tilde{x}=(-\vec{x},x_3)$ and momenta 
$\vec{p}=i\frac{\partial}{\partial \vec{x}}\to -i\frac{\partial}{\partial \vec{x}}=-\vec{p}$.
It is then clear that it exchanges A and B sites (sublattices) and Dirac points ($\vec{\rm K}_+$, $\vec{\rm K}_- = - \vec{\rm K}_+$),
respectively.
A convenient choice for the spinor transformation by spatial inversion is given by
\begin{equation}
\mathcal{P}: ~~\psi(x)\to \gamma_3 \psi(\tilde{x})~,~~\bar{\psi}(x)=\psi^\dagger(x)\gamma_3\to  \psi^\dagger(\tilde{x})~,
\end{equation}
or explicitely,
\begin{equation}
\psi(x) \to \left(
\begin{array}{c}
\psi^{B}_{K_-}(\tilde{x})\\
\psi^{A}_{K_-}(\tilde{x})\\
\psi^{A}_{K_+}(\tilde{x})\\
\psi^{B}_{K_+}(\tilde{x})
\end{array}
\right)~.
\end{equation}
For further application, let us also recall the well-known rule that Grassmann involution of a product of fields and a complex number $c$ gives, for example,
$$\left(c{\psi^{A}_{K_+}}^{\hspace{-5pt}*} \psi^{B}_{K_+}\right)^*=c^*{\psi^{B}_{K_+}}^{\hspace{-5pt}*} \psi^{A}_{K_+}
~~{\rm etc.},$$
with $c^*$ being the complex conjugate of $c$.

It is then, for example, straightforward to see that the current transforms as
$$\bar{\psi}\gamma_\mu \psi \to \bar{\psi}\widetilde{\gamma_\mu} \psi~,~~{\rm with} ~~
\widetilde{\gamma_\mu}=(-\vec{\gamma},\gamma_3)~,$$
as shown in table~\ref{tab:1}.

\subsubsection{Time reversal $\mathcal{T}$}

The antiunitary time reversal operation reverses momentum and spin, but not sublattices, and it interchanges $\vec{K_\pm}$
points \cite{I.29}. Accordingly, in euclidean notations we shall take 
\begin{equation}
\mathcal{T}: ~~\psi(x)\to (i\sigma_2)i\gamma_1\gamma_5 \psi({x})~,~~\bar{\psi}(x)\to - \bar{\psi}(x)i\gamma_1\gamma_5 (i\sigma_2)~,
\end{equation}
or explicitely,
\begin{equation}
\psi(x) \to (i\sigma_2)\left(
\begin{array}{c}
\psi^{A}_{K_-}({x})\\
\psi^{B}_{K_-}({x})\\
\psi^{B}_{K_+}({x})\\
\psi^{A}_{K_+}({x})
\end{array}
\right)~.
\end{equation}
Here $\sigma_2$ acts on the indices of physical spin contained in the suppressed flavor index $a$.
It is then straightforward to see that the current transforms, e.g., as 
$\bar{\psi} \gamma_\mu \psi \to \bar{\psi} \gamma_\mu \psi$ ~, see table~\ref{tab:1}. 
Antiunitary time reversal includes complex conjugation of c-numbers.
Note that in euclidean space $\mathcal{T}$ does not change the sign of the euclidean time $x_3$.
It also does not change the spatial components of the currents (see also table 1 of Ref.~\cite{Mesterhazy:2012ei}).

\subsubsection{Charge conjugation $\mathcal{C}$}

Charge conjugation exchanges particles and antiparticles (holes) leaving their spin and momentum unchanged.
This leads to the requirement that under  $\mathcal{C}$ the fermion current should transform as 
$\bar{\psi} \gamma_\mu \psi \to - \bar{\psi} \gamma_\mu \psi$
and the two Dirac points $\vec{K_\pm}$ should remain invariant.
The action of $\mathcal{C}$ on the four-spinor can then be taken as
\begin{equation}
\mathcal{C}: ~~\psi \to i\gamma_1\bar{\psi}^t~,~~\bar{\psi}\to (-i\gamma_1 \psi)^t~.
\end{equation}
One can see that $\mathcal{C}$ exchanges A and B sublattices. Indeed, 
\begin{equation}
\psi \to \left(
\begin{array}{c}
-{\psi^{B}_{K_+}}^{\hspace{-5pt}*}\\
-{\psi^{A}_{K_+}}^{\hspace{-5pt}*}\\
{\psi^{A}_{K_-}}^{\hspace{-5pt}*}\\
{\psi^{B}_{K_-}}^{\hspace{-5pt}*}
\end{array}
\right)~.
\end{equation}
Note that the charge conjugation matrix $C=i\gamma_1$ satisfies the relations
\begin{equation}
C=-C^{-1}=-C^\dagger=-C^t~,~~C^2=-1~,~~C\gamma_\mu^tC^{-1}= -\gamma_\mu~.
\end{equation}
With the above definitions, it indeed follows that  $\bar{\psi} \gamma_\mu \psi \to - \bar{\psi} \gamma_\mu \psi$~,
as quoted in table~\ref{tab:1}. 

\section{Stationarity/gap equations (vacuum case)}
\label{app:B}

From  the vacuum contribution to the thermodynamic potential in
Eq.~(\ref{3.29}) one obtains for the case of symmetric coupling
constants, $g_k = h_k = g$, the following stationarity/gap equations
\begin{eqnarray}
\label{B1}
\frac{\bar{\sigma}_1}{\bar{\rho}}\left(\frac{1}{g}\bar{\rho}+\frac{(\bar{\sigma}_2+\bar{\rho})^2}{2\pi}-{\rm
sign}(\bar{\sigma}_2-\bar{\rho})\frac{(\bar{\sigma}_2-\bar{\rho})^2}{2\pi}\right)=\kappa~,
\end{eqnarray}
\begin{eqnarray}
\label{B2}
\frac{\bar{\varphi}_i}{\bar{\rho}}\left(\frac{1}{g}\bar{\rho}+\frac{(\bar{\sigma}_2+\bar{\rho})^2}{2\pi}-{\rm
sign}(\bar{\sigma}_2-\bar{\rho})\frac{(\bar{\sigma}_2-\bar{\rho})^2}{2\pi}\right)=0~,
\end{eqnarray}
\begin{eqnarray}
\label{B3}
\left(\frac{1}{g}\bar{\sigma}_2+\frac{(\bar{\sigma}_2+\bar{\rho})^2}{2\pi}+{\rm
sign}(\bar{\sigma}_2-\bar{\rho})\frac{(\bar{\sigma}_2-\bar{\rho})^2}{2\pi}\right)=0~.
\end{eqnarray}
Clearly, for a nonvanishing condensate $\bar{\varphi}_i\neq 0$ ($i =1,2$), the bracket in Eq.~(\ref{B2}) should 
vanish which for $\kappa \neq 0$ is, however, impossible due to Eq.~(\ref{B1}). 
Thus, $\bar{\varphi}_i = 0$ and
$\bar{\rho}=\sqrt{\bar{\sigma}_1^2+\bar{\varphi}_1^2+\bar{\varphi}_2^2}=|\bar{\sigma}_1|$.
Analogously, $\bar{\sigma}_1=\bar{\sigma}_2$ cannot be a solution
of Eqs.~(\ref{B1}), (\ref{B3}). It is convenient to rewrite Eq.~(\ref{B1}) as ($\bar\sigma_1>0$)
\begin{eqnarray}
\label{B1a}
\frac{1}{g}\bar{\sigma}_1+\frac{\bar{\sigma}_1^2+\bar{\sigma}_2^2}{\pi}=\kappa~,~~(\bar{\sigma}_2<\bar{\sigma}_1)~,
\end{eqnarray}
\begin{eqnarray}
\label{B1b}
\frac{1}{g}\bar{\sigma}_1+\frac{2\bar{\sigma}_1\bar{\sigma}_2}{\pi}=\kappa~,~~(\bar{\sigma}_2>\bar{\sigma}_1)~,
\end{eqnarray}
and, analogously, for Eq.~(\ref{B3}),
\begin{eqnarray}
\label{B3a}
\frac{1}{g}\bar{\sigma}_2+\frac{\bar{\sigma}_1^2+\bar{\sigma}_2^2}{\pi}=0
~~~(\bar{\sigma}_2>\bar{\sigma}_1)~,
\end{eqnarray}
\begin{eqnarray}
\label{B3b}
\frac{1}{g}\bar{\sigma}_2+\frac{2\bar{\sigma}_1\bar{\sigma}_2}{\pi}=0
~~~(\bar{\sigma}_2<\bar{\sigma}_1)~.
\end{eqnarray}
Since $\Omega^{\rm ren}_{\rm vac}$ in Eq.~(\ref{3.29}) is an even function
of $\bar{\sigma}_2$, it is sufficient to study the global minimum
of $\Omega^{\rm ren}_{\rm vac}$ and the resulting gap equations only on
the set $\bar{\sigma}_{1,2}\geq 0$.
Next, suppose  $\bar{\sigma}_2\neq 0$. In this case, one obtains
from Eq.~(\ref{B3b}) for $0 < \bar{\sigma}_2 < \bar{\sigma}_1$,
$\bar{\sigma}_1=-\frac{\pi}{2g}$. 
Inserting this into Eq.~(\ref{B1a}) leads to
\begin{eqnarray}
-\frac{\pi}{2g^2}+\frac{\pi}{4g^2}+\frac{\bar{\sigma}_2^2}{\pi}=\kappa~,~~
\bar{\sigma}_2=\sqrt{\frac{\pi^2}{4g^2}+\pi \kappa}>\bar{\sigma}_1~, 
\label{B4}
\end{eqnarray}
which is in contradiction to our assumption $\bar{\sigma}_2\neq 0$
and $0<\bar{\sigma}_2 < \bar{\sigma}_1$. 
Thus, the solutions of the vacuum gap equations (\ref{B1}) - (\ref{B3}) are
$\bar{\varphi}_i=0$, $\bar{\sigma}_2 = 0$ and a finite $\bar{\sigma}_1$, given by
\begin{eqnarray}
\frac{1}{g}\bar{\sigma}_1+\frac{\bar{\sigma}_1^2}{\pi}=\kappa~,
~~\bar{\sigma}_1=-\frac{\pi}{2g} \stackrel{+}{(-)}
\sqrt{\left( \frac{\pi}{2g}\right)^2+\pi \kappa}= {\rm M}+\kappa |g|+O(\kappa^2)~.
 \label{B5}
\end{eqnarray}
Here we introduced the mass term ${\rm M} =\pi /|g|$ which, in the
chiral limit $\kappa= 0$, determines the dynamical fermion mass
$\bar{\sigma}_1 \equiv m(T=0, \mu= 0)$  at vanishing temperature
and chemical potential. 
One can see that the chosen positive solution in Eq.~(\ref{B5}) realizes the absolute minimum of
$\Omega^{\rm ren}_{\rm vac}$.

\section{Exciton polarization functions}
\label{app:C}

It is straightforward to show that 
in momentum-frequency space the polarization function $\Pi_E(\bar{q},i\nu_m)$ for excitons $E = \{ \sigma_i;\varphi_i\}$, 
shown in Fig.~\ref{fig:2}, is given by
\begin{eqnarray}
\Pi_E(\bar{q},i\nu_m)=-4\frac{N}{\beta} \sum_{n} \int
{\frac{d^2p}{(2\pi)^2}\frac{(i\omega_n+\mu)(i\omega_n+\mu-i\nu_m)-\vec{p}\cdot \vec{k}\pm m^2}
{\left[(i\omega_n+\mu)^2-E_p^2\right]\left[(i\omega_n+\mu-i\nu_m)^2-E_k^2\right]}}~,
 \label{C2}
\end{eqnarray}
where $\nu_m=2\pi m / {\beta}$ are bosonic Matsubara
frequencies, and the plus/minus sign in front of $m^2$ refers to
$\sigma_{1,2}/\varphi_{1,2}$ excitons.

For pedagogical reasons and a better understanding of the
important expression (\ref{4.10}), we will  now illustrate the
detailed calculation of the polarization function (\ref{C2}),
following the analogous procedure of Ref.~\cite{I.30}, which was
applied to $(3+1)$-dimensional models of quarks and composite
mesons. First, in order to perform the fermionic Matsubara sum
over $n$ in expression (\ref{C2}), one uses  standard partial
fraction decompositions and a compact notation with a sign
variable $s =\pm 1$ (a technique, already used in Ref.~\cite{I.31}). 
A typical expression reads
\begin{eqnarray}
\frac{1}{p_0^2-E^2_j}=\frac{1}{2E_j}\left(\frac{1}{p_0-E_j}-\frac{1}{p_0+E_j}\right)
=\sum_{s=\pm1}{\frac{1}{2E_j}\frac{s}{p_0-sE_j}}~.
 \label{C3}
\end{eqnarray}
Substituting $p_0 \rightarrow i\omega_n+\mu-i\nu_m$ and using $s^2=1$, one gets
\begin{eqnarray}
\sum_{s=\pm1}{\frac{1}{2E_j}\frac{s}{i\omega_n+\mu-i\nu_m-sE_j}}=\sum_{s=\pm1}{\frac{1}{2E_j}\frac{s}{i\omega_n-i\nu_m-s(E_j-s\mu)}}.
 \label{C4}
\end{eqnarray}
With this compact notation, the polarization function (\ref{C2}) is now written as
\begin{eqnarray}
\Pi_E(\vec{q},i\nu_m)=-4\frac{N}{\beta}\sum_{n}{\int{\frac{d^2p}{(2\pi)^2}}\sum_{s,s'}{\frac{ss'}{4E_p
E_k}\frac{(i\omega_n+\mu)(i\omega_n+\mu-i\nu_m)-\vec{p}\cdot\vec{k}\pm
m^2}{\left(i\omega_n-s\xi_p^{(-s)}\right)\left(i\omega_n-i\nu_m-s'\xi_k^{(-s')}\right)}}}~,
 \label{C5}
\end{eqnarray}
where
\begin{eqnarray}
\xi_j^{(-s)}=E_j-s\mu.
 \label{C6}
\end{eqnarray}
Generally, the Matsubara sum over fermion frequencies is performed by using the standard formulae \cite{I.32}
\begin{eqnarray}
\frac{1}{\beta}\sum_{n}{\frac{1}{i\omega_n-x}}=f(x)=\left(e^{\beta x}+1\right)^{-1}~,
 \label{C7}
\end{eqnarray}
\begin{eqnarray}
\frac{1}{\beta}\sum_{n}{\frac{1}{(i\omega_n-sE)}\cdot\frac{1}{(i\omega_n-i\nu_m-s'E')}}=\frac{f(s'E')-f(sE)}{i\nu_m+s'E'-sE}~.
 \label{C8}
\end{eqnarray}
Here Eq.~(\ref{C8}) simply follows by partial fraction decomposition
and then applying Eq.~(\ref{C7}). Note that the summation formula
Eq.~(\ref{C8}) cannot yet directly be applied to Eq.~(\ref{C5}), since
the numerator depends on $\omega_n$. 
To get rid of this, it is convenient to rewrite corresponding factors in the numerator by
subtracting and adding suitable terms. One thus has
\begin{eqnarray}
&&\frac{\left[\left(i\omega_n-s\xi_p^{(-s)}\right)+\left(s\xi_p^{(-s)}+\mu\right)\right]
\left[\left(i\omega_n - i\nu_m -
s'\xi_k^{(-s')}\right)+\left(s'\xi_k^{(-s')}+\mu\right)\right]}{\left(i\omega_n-s\xi_p^{(-s)}\right)\left(i\omega_n-i\nu_m-s'\xi_k^{(-s')}\right)}
\nonumber\\
&&=\left(1+\frac{s\xi_p^{(-s)}+\mu}{i\omega_n-s\xi_p^{(-s)}}\right)\left(1+\frac{s'\xi_k^{(-s')}+\mu}{i\omega_n-i\nu_m-s'\xi_k^{(-s')}}\right)
\nonumber\\
&&=\left[1+\frac{s\xi_p^{(-s)}+\mu}{i\omega_n-s\xi_p^{(-s)}}+\frac{s'\xi_k^{(-s')}+\mu}{i\omega_n-i\nu_m-s'\xi_k^{(-s')}}\right]+
\frac{\left(s\xi_p^{(-s)}+\mu\right)\left(s'\xi_k^{(-s')}+\mu\right)}{\left(i\omega_n-s\xi_p^{(-s)}\right)\left(i\omega_n-i\nu_m-s'\xi_k^{(-s')}\right)}~.
\nonumber\\
\label{C9}
\end{eqnarray}
Using Eqs.~(\ref{C5}), (\ref{C7}) and (\ref{C9}), one can see that the contribution of the square bracket term in the last line of
Eq.~(\ref{C9}) vanishes. 
Moreover, rewriting the factors in the numerator of the second term as  $(s\xi_p^{(-s)}+\mu) = s(E_p - s\mu) +\mu=s E_p$  etc., gives
\begin{eqnarray}
\Pi_E(\vec{q},i\nu_m)=-4\frac{N}{\beta}
\sum_{n}{\int\frac{d^2p}{(2\pi)^2}\sum_{s,s'}{\frac{ss'}{4E_p
E_k}\frac{sE_p s'E_k-\vec{p}\cdot \vec{k}\pm
m^2}{\left(i\omega_n-s\xi_p^{(-s)}\right)\left(i\omega_n-i\nu_m-s'\xi_k^{(-s')}\right)}}}~.
 \label{C10}
\end{eqnarray}
Finally, by applying now Eq.~(\ref{C8}) and taking into account that
$f(s'\xi_k^{(-s')} ) = f(s'E_k-\mu) \equiv f^-(s'E_k)$ etc., one obtains
\begin{eqnarray}
\Pi_E(\vec{q},i\nu_m)&=&-N\sum_{s,s'}\int {\frac{d^2p}{(2\pi)^2}\frac{f^-(s'E_k)-f^-(s
E_p)}{i\nu_m+s'E_k-sE_p} {\mathcal{T}}^{\mp}}~,
\nonumber\\
 \mathcal{T}^{\mp} &=&
%\left(
1-ss'\frac{\vec{p}\cdot \vec{k}\mp m^2}{E_p E_k}~,
%\right)
 \label{C11}
\end{eqnarray}
which is just the expression used in Eq.~(\ref{4.10}) of the main text. 
Note that in deriving Eq.~(\ref{C11}) use has been made of the fact that for bosonic Matsubara frequencies 
$\nu_m=2\pi m/\beta$, $f(x+\nu_m)=f(x)$. 
Let us remark that the compact notation of the excitonic polarization  function (\ref{C11}) reproduces the
general structure of the corresponding meson expression Eq.~({3.9}) of H\"ufner et al. in Ref.~\cite{I.17}.

For completeness, we also quote the Matsubara sum formula for bosons \cite{I.32}
\begin{eqnarray}
\frac{1}{\beta}\sum_{m}{\frac{1}{i\nu_m-x}}=-g(x)%=-(e^{\beta x}-1)^{-1}~.
=-\frac{1}{{\rm e}^{\beta x}-1}
 \label{C12}
\end{eqnarray}
Here  $g(x)$ is the Bose-Einstein distribution,% function, 
and $\nu_m $ are the bosonic Matsubara frequencies.

\end{document}